\documentclass[11pt]{mystyle}
\usepackage{times}

\usepackage{amssymb}
\usepackage{amsbsy}

\usepackage{amsmath}

\usepackage{psfig}
\psfigurepath{./figures/}
\pssilent

\newcommand{\ga}{\alpha}
\newcommand{\gb}{\beta}
\newcommand{\grg}{\gamma}
\newcommand{\gd}{\delta}

\newcommand{\gl}{\lambda}
\newcommand{\glh}{\mbox{$ \hat\lambda$}}

\newcommand{\gs}{\sigma}

\def\bm#1{\mbox{\boldmath $#1$}}

\newcommand{\vgrg}{\mbox{$\bm \gamma$}}

\newcommand{\vgl}{\mbox{$\bm \lambda$}}

\newcommand{\vpsi}{\mbox{$\bm \psi$}}

\newcommand{\mgG}{\mbox{$\bm \Gamma$}}
\newcommand{\mgGh}{\mbox{$\hat {\bm \Gamma}$}}
\newcommand{\mgD}{\mbox{$\bm \Delta$}}

\newcommand{\mgL}{\mbox{$\bm \Lambda$}}
\newcommand{\mgLh}{\mbox{$\bm {\hat \Lambda}$}}

\newcommand{\mLambda}{\mbox{$\bm \Lambda$}}

\newcommand{\bsigma}{{\bm\sigma}}

\newcommand{\rH}{^{ \raisebox{1pt}{$\rm \scriptscriptstyle H$}}}

\newcommand{\rE}{{\rm E}}

\newtheorem{theorem}{Theorem}[section]
\newtheorem{lemma}[theorem]{Lemma}

\newtheorem{prop}{Proposition}[section]
\newtheorem{claim}{Claim}[section]

\newtheorem{definition}{Definition}[section]
\newtheorem{question}{Question}[section]
\newtheorem{coro}{Corollary}[section]

\newcommand{\beq}{\begin{equation}}
\newcommand{\enq}{\end{equation}}

\newcommand{\bea}{\begin{array}}
\newcommand{\ena}{\end{array}}
\newcommand{\bds}{\begin {itemize}}
\newcommand{\eds}{\end {itemize}}
\newcommand{\bdf}{\begin{definition}}
\newcommand{\blm}{\begin{lemma}}
\newcommand{\edf}{\end{definition}}
\newcommand{\elm}{\end{lemma}}
\newcommand{\bthm}{\begin{theorem}}
\newcommand{\ethm}{\end{theorem}}
\newcommand{\bprp}{\begin{prop}}
\newcommand{\eprp}{\end{prop}}
\newcommand{\bcl}{\begin{claim}}
\newcommand{\ecl}{\end{claim}}
\newcommand{\bcr}{\begin{coro}}
\newcommand{\ecr}{\end{coro}}
\newcommand{\bquest}{\begin{question}}
\newcommand{\equest}{\end{question}}

\newcommand{\larrow}{{\larrow}}

\newcommand{\ie}{\hbox{i.e., }}
\newcommand{\eg}{\hbox{e.g., }}

\newcommand{\INR}{\ensuremath{\mathrm{INR}}}

\newcommand{\diag}{\ensuremath{\mathrm{diag}}}
\newcommand{\trace}{\ensuremath{\mathrm{tr}}}

\newcommand{\vect}{\ensuremath{\mathrm{vec}}}

\newcommand{\spann}{\ensuremath{\mathrm{span}}}

\newcommand{\argmin}{\ensuremath{\mathrm{arg}\min}}
\newcommand{\argmax}{\ensuremath{\mathrm{arg}\max}}

\newcommand{\tr}{\ensuremath{\mathrm{tr}}}

\newcommand{\cL}{\ensuremath{\mathcal{L}}}

\newcommand{\va}{{\ensuremath{\mathbf{a}}}}

\newcommand{\vb}{{\ensuremath{\mathbf{b}}}}

\newcommand{\vg}{{\ensuremath{\mathbf{g}}}}
\newcommand{\vgh}{{\ensuremath{\mathbf{\hat{g}}}}}
\newcommand{\vl}{{\ensuremath{\mathbf{l}}}}
\newcommand{\vm}{{\ensuremath{\mathbf{m}}}}
\newcommand{\vh}{{\ensuremath{\mathbf{h}}}}

\newcommand{\vr}{{\ensuremath{\mathbf{r}}}}
\newcommand{\vs}{{\ensuremath{\mathbf{s}}}}
\newcommand{\vsh}{{\ensuremath{\mathbf{\hat{s}}}}}

\newcommand{\vu}{{\ensuremath{\mathbf{u}}}}
\newcommand{\vv}{{\ensuremath{\mathbf{v}}}}

\newcommand{\vw}{{\ensuremath{\mathbf{w}}}}
\newcommand{\vwh}{{\ensuremath{\mathbf{\hat{w}}}}}

\newcommand{\vx}{{\ensuremath{\mathbf{x}}}}
\newcommand{\vxh}{{\ensuremath{\mathbf{\hat{x}}}}}

\newcommand{\ba}{{\ensuremath{\mathbf{a}}}}
\newcommand{\bah}{{\ensuremath{\mathbf{\hat{a}}}}}

\newcommand{\br}{{\ensuremath{\mathbf{r}}}}
\newcommand{\bs}{{\ensuremath{\mathbf{s}}}}
\newcommand{\bu}{{\ensuremath{\mathbf{u}}}}

\newcommand{\bw}{{\ensuremath{\mathbf{w}}}}
\newcommand{\bx}{{\ensuremath{\mathbf{x}}}}

\newcommand{\mA}{{\ensuremath{\mathbf{A}}}}

\newcommand{\mAt}{{\ensuremath{\mathbf{\tilde{A}}}}}
\newcommand{\mB}{{\ensuremath{\mathbf{B}}}}
\newcommand{\mBh}{{\ensuremath{\mathbf{\hat{B}}}}}
\newcommand{\mC}{{\ensuremath{\mathbf{C}}}}

\newcommand{\mF}{{\ensuremath{\mathbf{F}}}}

\newcommand{\mI}{{\ensuremath{\mathbf{I}}}}

\newcommand{\mJ}{{\ensuremath{\mathbf{J}}}}
\newcommand{\mL}{{\ensuremath{\mathbf{L}}}}
\newcommand{\mLb}{{\ensuremath{\mathbf{\bar{L}}}}}

\newcommand{\mP}{{\ensuremath{\mathbf{P}}}}

\newcommand{\mR}{{\ensuremath{\mathbf{R}}}}
\newcommand{\mRh}{{\ensuremath{\mathbf{\hat{R}}}}}
\newcommand{\mRt}{{\ensuremath{\mathbf{\tilde{R}}}}}

\newcommand{\mU}{{\ensuremath{\mathbf{U}}}}
\newcommand{\mUh}{{\ensuremath{\mathbf{\hat{U}}}}}

\newcommand{\mX}{{\ensuremath{\mathbf{X}}}}

\newcommand{\bP}{{\ensuremath{\mathbf{P}}}}

\newcommand{\bR}{{\ensuremath{\mathbf{R}}}}

\usepackage{latexsym}
\def\qed{{\unskip\nobreak\hfil\penalty50
    \hskip2em\hbox{}\nobreak\hfil{$\Box$}\parfillskip=0pt 
    \vspace{1ex plus 2pt}\par}}

\def\IC{\mathbb C}
\def\IN{\mathbb N}
\def\IZ{\mathbb Z}
\def\IR{\mathbb R}

\def\shat{^{\mathchoice{}{}%
 {\,\,\smash{\hbox{\lower4pt\hbox{$\widehat{\null}$}}}}%
 {\,\smash{\hbox{\lower3pt\hbox{$\hat{\null}$}}}}}}

\def\bSigma{{
      \ooalign{
      \smash{\hskip.4pt\raise.4pt\hbox{$\Sigma$}}\vphantom{}\crcr
      \smash{\hskip.7pt\raise.6pt\hbox{$\Sigma$}}\vphantom{}\crcr
      \smash{\hbox{$\Sigma$}}\vphantom{$\Sigma$}}
      \vphantom{\hbox{$\Sigma$}}
      }}
\def\bTheta{{
      \ooalign{
      \smash{\hskip.5pt\raise.5pt\hbox{$\Theta$}}\vphantom{}\crcr
      \smash{\hskip.0pt\raise.1pt\hbox{$\Theta$}}\vphantom{}\crcr
      \smash{\hbox{$\Theta$}}\vphantom{$\Theta$}}
      \vphantom{\hbox{$\Theta$}}
      }}
\def\bDelta{{
      \ooalign{
      \smash{\hskip.4pt\raise.4pt\hbox{$\Delta$}}\vphantom{}\crcr
      \smash{\hskip.7pt\raise.6pt\hbox{$\Delta$}}\vphantom{}\crcr
      \smash{\hbox{$\Delta$}}\vphantom{$\Delta$}}
      \vphantom{\hbox{$\Delta$}}
      }}

\makeatletter

\def\bordermatrix#1{\begingroup \m@th
  \@tempdima 8.75\p@
  \setbox\z@\vbox{%
    \def\cr{\crcr\noalign{\kern2\p@\global\let\cr\endline}}%
    \ialign{$##$\hfil\kern2\p@\kern\@tempdima&\thinspace\hfil$##$\hfil
      &&\quad\hfil$##$\hfil\crcr
      \omit\strut\hfil\crcr\noalign{\kern-\baselineskip}%
      #1\crcr\omit\strut\cr}}%
  \setbox\tw@\vbox{\unvcopy\z@\global\setbox\@ne\lastbox}%
  \setbox\tw@\hbox{\unhbox\@ne\unskip\global\setbox\@ne\lastbox}%
  \setbox\tw@\hbox{$\kern\wd\@ne\kern-\@tempdima\left[\kern-\wd\@ne
    \global\setbox\@ne\vbox{\box\@ne\kern2\p@}%
    \vcenter{\kern-\ht\@ne\unvbox\z@\kern-\baselineskip}\,\right]$}%
  \null\;\vbox{\kern\ht\@ne\box\tw@}\endgroup}
\makeatother

\newcommand{\DL}{\begin{dashlist}}
\newcommand{\DLE}{\end{dashlist}}
\newcommand{\vgc}{{\ensuremath{\mathbf{\bar{g}}}}}
\newcommand{\mXh}{{\ensuremath{\mathbf{\hat{X}}}}}

\makeatletter
\def\argmin{\mathop{\operator@font arg\,min}}
\def\argmax{\mathop{\operator@font arg\,max}}
\makeatother

\newcommand{\SI}{\begin{indlist}}
\newcommand{\EI}{\end{indlist}}

\newcommand{\vuh}{\ensuremath{\mathbf{\hat{u}}}}
\def\INR{\mbox{INR}}

\title{Radio astronomical imaging in the presence of strong radio interference}

\author{Amir Leshem\thanks{Amir Leshem was supported by the NOEMI project 
    of the STW under contract no.~DEL77-4476. Parts of this research have
been presented in the workshop ``Perspectives in Radio-Astronomy: Technologies for large arrays, Dwingeloo, Netherlands 1999''.} \  and Alle-Jan van der Veen \\
    Dept. of Electrical Engineering\\
    Delft University of Technology\\
    2628 CD Delft, The Netherlands\\
    Email: leshem,allejan@cas.et.tudelft.nl
}
\begin{document}
\maketitle

\abstract{
    Radio-astronomical observations are increasingly contaminated by
    interference, and suppression techniques become essential. A powerful
    candidate for interference mitigation is adaptive spatial filtering.
    We study the effect of spatial filtering techniques on radio
    astronomical imaging.  Current deconvolution procedures such as
    CLEAN are shown to be unsuitable to spatially filtered data, and
    the necessary corrections are derived.  To that end, we reformulate
    the imaging (deconvolution/calibration) process as a sequential
    estimation of the locations of astronomical sources.
    This not only leads to an extended CLEAN algorithm, the formulation 
    also allows to insert other array signal processing techniques for
    direction finding, and gives estimates of the expected 
    image quality and the amount of interference suppression that can be
    achieved.
    Finally, a maximum likelihood procedure for the imaging is derived,
    and an approximate ML image formation technique is proposed to
    overcome the computational burden involved.
    Some of the effects of the new algorithms are shown in simulated images.
}


\section{Introduction}
The future of radio astronomical discoveries depends on achieving better
resolution and sensitivity while maintaining immunity to terrestrial
interference which is rapidly growing. The last two demands are obviously
contradicting as improved sensitivity implies receiving more interfering
signals. One possible track is to switch to massive phased array
technology.  If instead of the huge dishes which became
the trademark of radio astronomy, we use phased array radio-telescopes
comprised of tens of thousands of small elements, then we gain both
in terms of resolution and sensitivity while increasing the flexibility
to mitigate
interference. The international effort in this direction is coordinated
under the framework of the Square Kilometer Array project (SKA).  In this
paper we try to analyze the effect of on-line interference rejection on
the image formation process in such an instrument.

We briefly describe the current status of radio astronomical imaging;
for a more extensive overview the reader is referred to \cite{thompson86}
\cite{yen85} or \cite{perley89}. The principle of radio interferometry has
been used in radio astronomy since 1946 when Ryle and Vonberg constructed
a radio interferometer using  dipole antenna arrays \cite{ryle52}. During
the 1950's several radio interferometers which use the synthetic aperture
created by movable antennas have been constructed. In 1962 the
principle of aperture synthesis using earth rotation has been proposed
\cite{ryle62}. The basic idea is to exploit the rotation of the earth to
obtain denser coverage of the visibility domain (spatial Fourier domain).

The first instrument to use this principle was the five kilometer
Cambridge radio telescope.  During the 1970's new instruments with
large aperture have been constructed. Among these we find the Westerbork
Synthesis Radio Telescope (WSRT) in the Netherlands and the Very Large
Array (VLA) in the USA.  Even these instruments subsample the Fourier
domain, so that unique reconstruction is not possible without some
further processing known as deconvolution. The deconvolution process
uses some a-priori knowledge about the image to remove the effect of
``dirty beam'' side-lobes.

Two principles dominate the astronomical imaging deconvolution.
The first method was proposed by Hogbom \cite{hogbom74} and is known
as CLEAN. The CLEAN method is basically a sequential Least-Squares (LS)
fitting procedure in which the brightest source location and power are
estimated. The response of this source is removed from the image and then
the process continues to find the next brightest source, until the residual
image is noise-like.  During the years it has been partially
analyzed \cite{schwarz78}, \cite{schwarz79} and \cite{tan86}. However
full analysis  of the method is still lacking due to its iterative nature.

A second root proposed by Jaynes \cite{Jaynes57}, \cite{jaynes70} is
maximum entropy deconvolution (MEM). The basic idea behind MEM is
the following.  Among all images which are consistent with the measured
data and the noise distribution not all satisfy the positivity demand,
i.e., the sky brightness is a positive function. Consider only those
that satisfy the positivity demand.
  From these select the one that is most likely to have been created randomly.
This idea has also been proposed in \cite{frieden72} and applied to
radio astronomical imaging in \cite{gull78}. Other approaches based
on the differential entropy have also been proposed \cite{ables74} and
\cite{wernecke77}. An extensive collection of papers discussing the various 
methods and aspects of maximum entropy can be found in the various papers in 
\cite{roberts84}.

The above mentioned algorithms assume perfect knowledge of the
instrumental response (point spread function). Due to various internal
and external effects this assumption holds only approximately. One way
to overcome this problem is the use of calibrating sources. An unresolved
source with known parameters is measured, and by relating the model errors
to the array elements a set of calibration equations is solved. A much
more appealing solution is to try to improve the fitting between the
data and the sky model by adjusting the calibration parameters. Another
possibility \cite{noordam82} is to use the redundant structure of the
array to solve for the calibration parameters (this is possible only
for some arrays which have redundant baselines, such as the WSRT). A good
overview of the various techniques is given in \cite{pearson84}.

A major problem facing radio astronomy is the accelerated use of the
electro-magnetic spectrum. Even in bands which are reserved to radio
astronomical observation one can find interference, e.g., sidelobes
of emissions from the Iridium or GLONASS satellites. As was shown
in \cite{thompson82} interferometric arrays are less sensitive to
interference than single dish instruments. However the interference
still appears in images, especially for observations at frequency bands
not specifically reserved for radio astronomy. Many efforts all over
the world are currently put into improving the interference mitigation
capabilities of radio telescopes.  The methods considered span over
all possible domains: temporal \cite{weber97}, \cite{friedman96a},
\cite{bradely98}, spatio-temporal  \cite{kasper82}, spatio-spectral
\cite{leshem99spawc}, \cite{leshem99ahos}, \cite{leshem98t1}, and wavelet
based methods \cite{maslakovic96}. Also considered are techniques which
use statistical and deterministic properties such as non-Gaussianity and
constant modulus  of the interferers in order to use blind-beamforming,
to remove the interferers.

As far as multi-element synthesis imaging radio telescopes are
concerned, the spatial methods are extremely appealing, since each
interferer has its own ``spatial signature''. Estimating these signatures
enables efficient mitigation of the interferer using phased-array
techniques. However two important questions shade over these methods. As
we will show in this paper, applying time varying spatial filters
makes the point spread function space varying. In particular the image
formation techniques will have to be modified accordingly. To cope with
this problem we reformulate the classical Fourier imaging framework  in
a parametric manner more appropriate for statistical analysis of the
interference mitigation. This enables us to incorporate the spatial
filtering naturally into the image formation process.

Our reformulation of the image formation problem describes the
measurements as a set of covariance matrices, measured at the various
observation epochs.  This yields a model where the array response is time
varying. Previous research on time varying arrays and their application
to direction-of arrival (DOA) estimation includes \cite{zeira95}, where
a maximum likelihood estimator for a single source and the corresponding
Cramer-Rao Bound (CRB) are derived. For multiple sources \cite{zeira96}
proposes three approaches: A modified beamforming, a virtual interpolated
array and focusing matrices. The main drawback of the last two methods
is that they do not lend themselves to estimating more sources than the
number of physical sensors. In \cite{sheinvald98} a generalized LS (GLS)
approach is proposed. The idea is to present the problem of estimating
the source powers (given the DOA parameters) as a linear problem. This
enables a closed form solution for the powers. Substituting back into
the GLS estimators, the problem is reduced to a multidimensional search
problem over the DOA's. The main drawback of this approach is the need to
invert very large matrices ($p^2 \times p^2$, where $p$ is the number of
physical sensors). It is also shown that the asymptotic performance of the
GLS and the maximum likelihood estimators is identical up to second order,
which makes it very attractive. All these papers deal with the case of a
fully known array response, and are limited to one-dimensional location
parameters. 
In this paper we also extend the above methods in several directions. First
we use the eigenstructure, to mitigate strong interferers for which
we do not know the array response, and analyze the efficiency of this
procedure. Second we propose other relatively ``low'' complexity
approaches, based on eigenstructure and Minimum Variance Distortionless
Response (MVDR). Finally we discuss a maximum likelihood approach to the 
astronomical image formation problem, derive an approximate ML estimator 
(AML) of low complexity 
and discuss the relation of the AML to the CLEAN algorithm. Our treatment of 
the radio-astronomical imaging can be applied to other problems of imaging 
with time varying sensor responses such as ISAR/SAR radar imaging in the 
presence of strong jammers. The model then generalizes the model given in 
\cite{snyder89} to the time varying context. For the case of an array with 
constant time behavior (connected element array) methods of computing the MLE 
have been proposed in \cite{snyder89} using the EM method or \cite{fessler94} 
using the SAGE algorithm. However, there is no straightforward extension of 
these methods to our context. Although an EM algorithm can be applied, the
dimensionality of the parameter space makes it infeasible to perform
full MLE without very good initialization. The paper \cite{osullivan98}
contains an extensive overview on image formation principles and
algorithms, as well as previous work on parametric image formation in
other fields such as radar and tomography.

To allow the paper to be of use both to the information theory/signal
processing and to the radio astronomical communities the introductory part
is of a tutorial nature.
The structure of the paper is as follows.  In section
\ref{sec:astron} we describe the astronomical measurement process and
introduce an often-used coordinate system.  The measurement equation
is subsequently rephrased in a more convenient matrix formulation in
section \ref{sec:array}, and extended with the effect of interference.
In section \ref{sec:spatfilt} we describe several basic spatial
filtering approaches to on-line interference suppression, and compute
the residual interference after adaptive estimation of its parameters
for one specific case.  In section \ref{sec:imag} and \ref{sec:bfimag} we
discuss the image formation process, first based on classical techniques
(CLEAN), then extended to other beamforming methods and taking the spatial
filtering into account.  Finally, we derive an approximate ML algorithm
for image formation.  We end up with conclusions regarding future
implementation of on-line interference suppression in radio-astronomy.

\section{Astronomical measurement equations}		\label{sec:astron}

    In this section we describe a simplified mathematical model for the 
    astronomical measurement and imaging process.  Our discussion follows
    the introduction in \cite{perley89}.
    We begin with the measurement equation, to
    reformulate it into a matrix form in the next section.
    This will allow us to obtain a uniform
    description of various astronomical imaging operations such as
    deconvolution and self-calibration.

\begin{figure}
  \begin{center}
\begin{picture}(0,0)%
\includegraphics{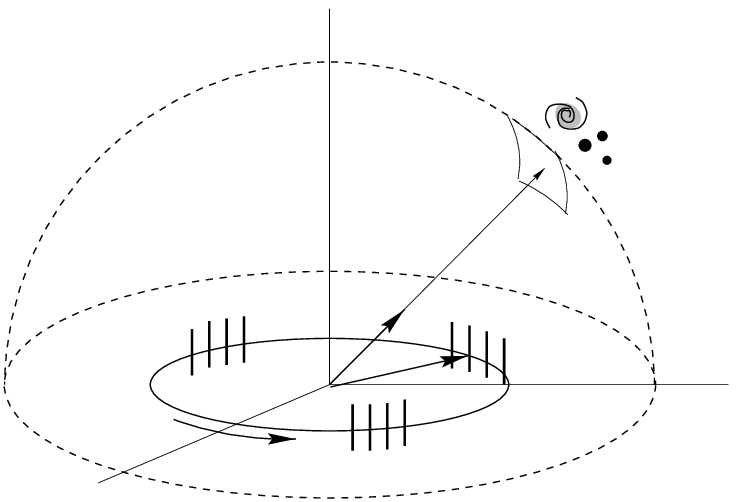}%
\end{picture}%
\setlength{\unitlength}{1460sp}%
\begingroup\makeatletter\ifx\SetFigFont\undefined
\def\x#1#2#3#4#5#6#7\relax{\def\x{#1#2#3#4#5#6}}%
\expandafter\x\fmtname xxxxxx\relax \def\y{splain}%
\ifx\x\y   
\gdef\SetFigFont#1#2#3{%
  \ifnum #1<17\tiny\else \ifnum #1<20\small\else
  \ifnum #1<24\normalsize\else \ifnum #1<29\large\else
  \ifnum #1<34\Large\else \ifnum #1<41\LARGE\else
     \huge\fi\fi\fi\fi\fi\fi
  \csname #3\endcsname}%
\else
\gdef\SetFigFont#1#2#3{\begingroup
  \count@#1\relax \ifnum 25<\count@\count@25\fi
  \def\x{\endgroup\@setsize\SetFigFont{#2pt}}%
  \expandafter\x
    \csname \romannumeral\the\count@ pt\expandafter\endcsname
    \csname @\romannumeral\the\count@ pt\endcsname
  \csname #3\endcsname}%
\fi
\fi\endgroup
\begin{picture}(9422,6371)(266,-6645)
\put(5581,-6046){\makebox(0,0)[lb]{\smash{\SetFigFont{9}{10.8}{rm}E}}}
\put(5356,-4111){\makebox(0,0)[rb]{\smash{\SetFigFont{9}{10.8}{rm}$\bs$}}}
\put(6871,-2701){\makebox(0,0)[rb]{\smash{\SetFigFont{9}{10.8}{rm}$\bR$}}}
\put(5596,-4921){\makebox(0,0)[rb]{\smash{\SetFigFont{9}{10.8}{rm}$\br$}}}
\put(4741,-6076){\makebox(0,0)[rb]{\smash{\SetFigFont{9}{10.8}{rm}W}}}
\end{picture}

\begin{picture}(0,0)%
\includegraphics{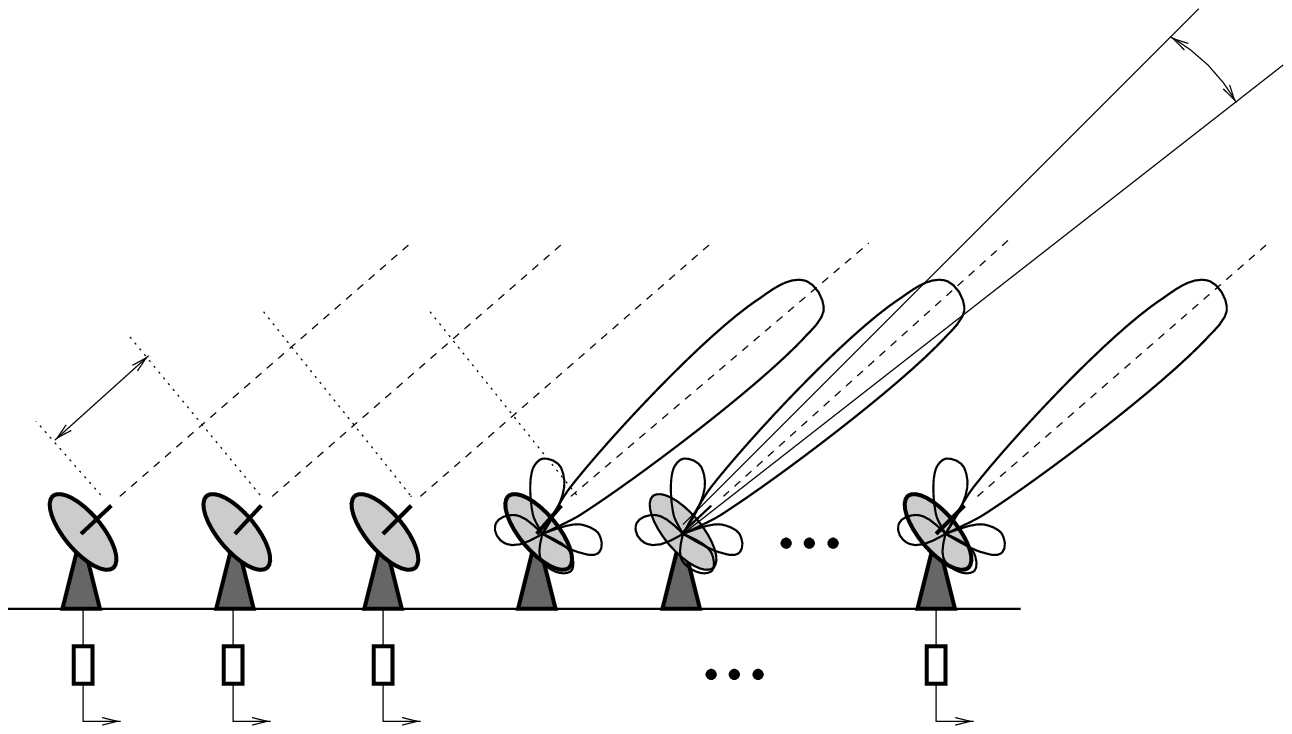}%
\end{picture}%
\setlength{\unitlength}{2368sp}%
\begingroup\makeatletter\ifx\SetFigFont\undefined
\def\x#1#2#3#4#5#6#7\relax{\def\x{#1#2#3#4#5#6}}%
\expandafter\x\fmtname xxxxxx\relax \def\y{splain}%
\ifx\x\y   
\gdef\SetFigFont#1#2#3{%
  \ifnum #1<17\tiny\else \ifnum #1<20\small\else
  \ifnum #1<24\normalsize\else \ifnum #1<29\large\else
  \ifnum #1<34\Large\else \ifnum #1<41\LARGE\else
     \huge\fi\fi\fi\fi\fi\fi
  \csname #3\endcsname}%
\else
\gdef\SetFigFont#1#2#3{\begingroup
  \count@#1\relax \ifnum 25<\count@\count@25\fi
  \def\x{\endgroup\@setsize\SetFigFont{#2pt}}%
  \expandafter\x
    \csname \romannumeral\the\count@ pt\expandafter\endcsname
    \csname @\romannumeral\the\count@ pt\endcsname
  \csname #3\endcsname}%
\fi
\fi\endgroup
\begin{picture}(10553,5832)(560,-6481)
\put(1876,-6436){\makebox(0,0)[lb]{\smash{\SetFigFont{10}{12.0}{rm}$x_1(t)$}}}
\put(3076,-6436){\makebox(0,0)[lb]{\smash{\SetFigFont{10}{12.0}{rm}$x_2(t)$}}}
\put(4276,-6436){\makebox(0,0)[lb]{\smash{\SetFigFont{10}{12.0}{rm}$x_3(t)$}}}
\put(8701,-6436){\makebox(0,0)[lb]{\smash{\SetFigFont{10}{12.0}{rm}$x_{14}(t)$}}}
\put(1651,-6061){\makebox(0,0)[lb]{\smash{\SetFigFont{10}{12.0}{rm}$T_1$}}}
\put(2851,-6061){\makebox(0,0)[lb]{\smash{\SetFigFont{10}{12.0}{rm}$T_2$}}}
\put(4051,-6061){\makebox(0,0)[lb]{\smash{\SetFigFont{10}{12.0}{rm}$T_3$}}}
\put(8476,-6061){\makebox(0,0)[lb]{\smash{\SetFigFont{10}{12.0}{rm}$T_{14}$}}}
\put(1576,-3361){\makebox(0,0)[rb]{\smash{\SetFigFont{10}{12.0}{rm}geometric}}}
\put(1576,-3646){\makebox(0,0)[rb]{\smash{\SetFigFont{10}{12.0}{rm}delay}}}
\end{picture}

  \end{center}
  \caption{$(a)$ The emitted electrical field from the celestial sphere 
  is received by a rotating telescope array; 
  $(b)$ geometrical delay compensation}
  \label{Fig:space}
\end{figure}

    The signals received from the celestial sphere may be considered as
    spatially incoherent wideband random noise. It is
    possibly polarized and perhaps contains spectral absorption or
    emission lines.
    Rather than considering the emitted electric field at a location on
    the celestial sphere, astronomers try to recover the {\em
    intensity} (or brightness) $I_f(\vs)$ in the direction of
    unit-length vectors $\vs$, where $f$ is a specific frequency.
    Let $E_f(\vr)$ be the received celestial electric field at a location
    $\vr$ on earth (see figure \ref{Fig:space}$(a)$). 
    The measured correlation of the electric fields
    between two identical sensors $i$ and $j$ with locations $\vr_i$ and $\vr_j$
    is called a {\em visibility} and is (approximately) given by
    \cite{perley89}\footnote{To simplify notation we do not include in our model the directional response of the elements of the radio telescope. This can be included in a straightforward manner like in \cite{perley89} chapter 1 section 4.3}
\[
    V_f(\vr_i,\vr_j)
    :=    \rE\{ E_f(\vr_i) \overline{E_f(\vr_j)} \}
    \;=\; \int_{\mbox{sky}}
          I_f(\vs) e^{-2\pi \jmath f\; \bs^T(\vr_i - \vr_j)/c}\; d\Omega
    \,.
\]
    ($\rE\{\,\cdot\,\}$ is the mathematical expectation operator,
    the superscript $^T$ denotes the transpose of a vector, and overbar denotes
    the complex conjugate.)
    Note that it is only dependent on the oriented distance $\vr_i-\vr_j$
    between the two antennas; this vector is called a baseline.

    For simplification, we may sometimes
    assume that the astronomical sky is a collection of $d$
    discrete point sources (maybe unresolved). This gives
\[
    I_f(\vs) = \sum_{l=1}^d I_f(\vs_l) \delta(\vs - \vs_l) \,,
\]
    where $\vs_l$ is the coordinate of the $l$'th source, and thus
\begin{equation}				\label{eq:discretesource}
    V_f(\vr_i,\vr_j) \;=\; 
	  \sum_{l=1}^d 
	    \,
	    I_f(\vs_l)
	    \,
	    e^{-2\pi \jmath f\, \vs_l^T(\vr_i - \vr_j)/c}  \,.
\end{equation}

    Up to this point we have worked in an arbitrary coordinate system.
    For earth rotation synthesis arrays, a coordinate system is often
    introduced as follows.  We assume an array with antennas that
    have a small field of view and that track a reference source
    direction $\bs_0$ in the sky.
    Other locations in the field of view can be written as
$
    \bs = \bs_0 + \bsigma\,, \bs_0 \perp \bsigma
$
    (valid for small $\bsigma$) and a natural coordinate system is
\[
       \bs_0 = [0,\, 0,\, 1]^T 
       \,,
       \qquad
       \bsigma = [\ell,\, m,\, 0]^T \,.
\]
    Similarly, for a planar array, the receiver baselines can be
    parameterized as
\[
    \br_i - \br_j = \lambda [u,\, v,\, w]^T \,,
    \qquad \lambda = \displaystyle\frac{c}{f} \,.
\]
    The measurement equation in $(u,v,w)$ coordinates thus becomes
\begin{equation} 					\label{eq:fourier1}
    V_f(u,v,w)
    \;=\; e^{-2\pi \jmath w} 	  
	  \int\!\!\!\int\; 
	  I_f(\ell,m)\, e^{-2 \pi \jmath(u\ell + v m)}\, d\ell d m
    \,.
\end{equation}
    The factor $e^{-2\pi \jmath w}$ is caused by the {\em geometrical delay}
    associated to the reference location, and can be compensated by
    introducing a slowly time-variant delay (see figure \ref{Fig:space}$(b)$).
    This synchronizes the center of the field-of-view and makes the
    reference source location appear as if it was at the north pole.
    After compensation, we arrive at a measurement equation in $(u,v)$
    coordinates only, 
\begin{equation}					\label{eq:fourier}
    V_f(u,v)
    \;=\; 
	  \int\!\!\!\int\; 
	  I_f(\ell,m)\, e^{-2 \pi \jmath(u\ell + v m)}\, d\ell d m
    \,.
\end{equation}
    It has the form of a Fourier transformation. 

    The function $V_f(u,v)$ is sampled at various coordinates $(u,v)$
    by first of all taking all possible sensor pairs $i,j$ or baselines
    $\br_i-\br_j$, and second by realizing that the sensor locations
    $\br_i$, $\br_j$ are actually time-varying since the earth
    rotates.  Given a sufficient number of samples in the $(u,v)$ domain, the
    relation can be inverted to obtain an image (the `map'),
    which is the topic of section \ref{sec:imag}.

\section{Array signal processing formulation}		\label{sec:array}

\subsection{Obtaining the measurements}

    We will now describe the situation from an array signal processing
    point of view.  The signals received by the antennas are
    amplified and downconverted to baseband.  A time-varying delay for
    every antenna is also introduced, to compensate for the
    geometrical delay.
    Following traditional array signal processing practices, the
    signals at this point are called $x_i(t)$ rather than $E_f(\br)$,
    and are stacked in vectors 
\[
    \bx(t) = \left[x_1(t),\ldots,  x_{p}(t) \right]^T \,,
\]
    where $p$ is the number of antennas.  These are then processed by
    a correlation stage.

    It will be convenient to assume that $\bx(t)$ is first split by a
    bank of narrow-band sub-band filters into a collection of
    frequency-components $\bx_f(t)$.
    The main output of the telescope hardware is then a sequence of
    empirical correlation matrices $\mRh_f(t)$ of cross-correlations of 
    $\bx_f(t)$, for a set of frequencies
    $f \in \{f_k\}$ covering a 10 MHz band or so, and for a set of times
    $t \in \{t_k\}$ covering up to 12 hours 
\footnote{
    Many telescope sites including WSRT follow actually a
    different scheme where the signals are first correlated at several
    lags and subsequently Fourier transformed.  This leads to similar
    results.}.
  Each correlation matrix
    $\mRh_f(t)$ is an estimate of the true covariance matrix $\mR_f(t)$,
\begin{equation}					\label{eq:avgNT}
    \mR_f(t) \;=\; {\rm E} \{ \bx_f(t) \bx_f(t)\rH \} \,,
    \qquad
    \mRh_f(t) = 
    \frac{1}{N}\sum_{n=0}^{N-1} \bx_f(t + nT) \bx_f(t + nT)\rH \,,
\end{equation}
    where the superscript $\rH$ denotes a complex conjugate transpose,
    $T$ is the sample period of $\bx_f(t)$ and $N$ is the number of
    samples over which is averaged. 
    The matrices $\mRh_f(t)$ are stored for off-line spectral analysis and
    imaging.
    Typically, each sub-band has a bandwidth in the order of 100~kHz or
    less.  Due to the sub-band filtering, the original sampling rate of
    $\bx(t)$ is reduced accordingly, resulting in $T$ in the order of
    $10\ \mu$s and the number of samples $N$ in the order of $10^5$ for each 
    sub-band. $f$ represents the center frequency in a sub-band. From now on 
    we consider the sub-bands independently ignoring that they are really 
    connected.  Consequently, in future equations we drop the dependence
    on $f$ in the notation.

    The connection of the correlation matrices $\mR(t)$ to the 
    visibilities $V(u,v)$ in section \ref{sec:astron} is as follows.
    Each entry $r_{ij}(t)$ of the matrix $\mR(t)$ is a sample of this
    visibility function for a specific coordinate $(u,v)$ corresponding
    to the baseline vector 
$\br_i(t) - \br_j(t) = \lambda [u_{ij}(t),\; v_{ij}(t),\; w_{ij}(t)]$ 
    between telescopes $i$ and $j$ at time $t$:
\begin{equation}					\label{eq:visdef}
    V(u_{ij}(t),v_{ij}(t)) \equiv r_{ij}(t) \,.
\end{equation}
    Note that we can obtain only a discrete set of $(u,v)$ sample points.
    Indeed, the number of instantaneous independent baselines between
    $p$ antennas is at most $\frac{1}{2}p(p-1)$. Also, using the
    earth rotation, we have a finite set $\{t_k,\; k=1, \cdots, K\}$,
    where the number of epochs $K$ is given by the ratio of the observation
    time and the covariance averaging time (\eg $K =$ 12 h/30 sec =
    1440 samples). 
    The available
    sample coordinates $\{u_{ij,k},v_{ij,k}\}$ give an irregular cover of the
    $(u,v)$ plane.  For an East-West line array such as WSRT, the points
    lie on ellipses.
    A practical issue is the implementation of the geometrical delay 
    compensation. It is usually introduced only at the 
    back end of the receiver.
    At this point, also a phase correction
    is needed to compensate for the factor $e^{-2 \pi \jmath  w_{ij}(t)}$ 
    in the measurement equation (\ref{eq:fourier1}).
    This is referred to as {\em fringe correction} \cite{thompson86}.
    Since the earth rotates, $w_{ij}(t)$ is slowly time varying, with a
    rate of change in the order of 0-10 Hz depending on source declination
    and baseline length.

\subsection{Matrix formulation}

    For the discrete source model, we
    can now formulate our measurement equations in terms of matrices.
    Let $\vr_0(t_k)$ be an arbitrary and time-varying 
    reference point, typically at one of the elements
    of the array, and let us take the $(u,v,w)$ coordinates of the other
    telescopes with respect to this reference,
$
   \br_i(t) - \br_0(t) \;=\; \lambda [u_{i0}(t),\; v_{i0}(t),\; w_{i0}(t)] \,,
   \qquad
   i = 1, \cdots, p\,.
$
    Equation (\ref{eq:discretesource}) can then be
    written slightly differently as
\[
\bea{rcl}
    V(\vr_i(t),\vr_j(t)) &=&
	  \displaystyle\sum_{l=1}^d 
	    e^{-2\pi \jmath f\, \vs_l^T(\vr_i - \vr_0)/c} 
	    \,
	    I(\vs_l)
	    \,
	    e^{2\pi \jmath f\, \vs_l^T(\vr_j - \vr_0)/c} 
    \\
    \Leftrightarrow
    \quad
    V(u_{ij}(t),v_{ij}(t)) &=& 
	  \displaystyle\sum_{l=1}^d  
	    e^{-2 \pi \jmath(u_{i0}(t) \ell_l + v_{i0}(t) m_l)}
	    \cdot
	    I(\ell_l,m_l)
	    \cdot
	    e^{2 \pi \jmath(u_{j0}(t) \ell_l + v_{j0}(t) m_l)} \,.
\ena
\]
    In terms of correlation matrices, this equation can be written as 
   $\mR_k \;=\; \mA_k  \mB \mA\rH_k$,
    where $\mR_k \equiv \mR(t_k)$, 
\[
\mA_k = \left[\va_k(\ell_1,m_1),\ldots,\va_k(\ell_d,m_d)\right] 
\]
and
\begin{eqnarray}
\label{eq:vadef} 
    \va_k(\ell,m) &=&
    \left[ \bea{c}
	e^{-2\pi \jmath(u_{10}(t_k) \ell + v_{10}(t_k) m)} \\
	\vdots \\
	e^{-2\pi \jmath(u_{p0}(t_k) \ell + v_{p0}(t_k) m)}
    \ena \right] \\
\nonumber
    \mB &=&
    \left[ \bea{ccc}
	I(\ell_1,m_1) & & \mbox{\bf 0}\\
	& \ddots & \\
	\mbox{\bf 0} & & I(\ell_d,m_d)
    \ena \right] 
\end{eqnarray}
    The vector function $\va_k(\ell,m)$ is called
    the {\em array response vector} in array signal processing.  It
    describes the response of the telescope array
    to a source in the direction 
    $(\ell,m)$. As usual, the array response is frequency dependent.
    In this case, the response is also slowly time-varying due to the
    earth rotation.  Note, very importantly, that the function as shown
    here is completely known.

    More realistically, the array response is less perfect.
    An important effect is that each antenna may have a different complex
    receiver gain, $\gamma_i(t)$, dependent on many angle-independent
    effects such as cable losses, amplifier gains, and (slowly) varying
    atmospheric conditions.  
    We also have to realize that
    most of the received signal consists of additive system noise.
    When this noise is zero mean, independent among
    the antennas (thus spatially white), and identically distributed, then
    it has a covariance matrix that is a multiple of the identity matrix,
    $\sigma^2 \mI$, where $\sigma^2$ is the noise power on a single antenna
    inside the subband which we consider.  Usually the noise is assumed to be 
    Gaussian. 
    The resulting model of the received covariance matrix then becomes
\begin{equation} 					\label{self_cal_model}
    \mR_k \;=\; \mgG_k \mA_k \mB \mA\rH_k \mgG_k\rH  \;+\;  \gs^2 \mI
\end{equation}
    where
\begin{equation}
\label{def_Gammak}
    \mgG_k =
    \left[ \bea{ccc}
	\gamma_{1,k} & & \mbox{\bf 0}\\
	 & \ddots & \\
	\mbox{\bf 0}& & \gamma_{p,k}
    \ena \right] \,.
\end{equation}
    Note that this assumes that the noise is introduced {\em after} the 
    varying receiver gains.
    This assumption is reasonable if the channels from the 
    low-noise amplifier (LNA) outputs to the analog-to-digital converter (ADC)
    units are equal. Otherwise, it is still reasonable to assume
    that the noise is spatially white, i.e., the noise covariance matrix is
    diagonal.  We can assume that the receivers noise power can be estimated 
    using various calibration techniques; a simple diagonal scaling will then
    bring us back to the model (\ref{self_cal_model}).

\subsection{RF interference} 					\label{sec:RFI}
    Radio frequency interference (RFI) usually enters the antennas through the
    sidelobes of the main beam.  It can be stronger or weaker than the
    system noise.  An important property is that it has a certain {\em
    directivity}, so that it does not average out in the correlation
    process.  Examples of harmful RFI are television broadcasts,
    geolocation satellites (GPS, GLONASS), taxi dispatch systems,
    airplane communication and navigation signals, wireless mobile
    communication (GSM) and satellite communication signals (Iridium).
    Thus, interferers may be continuous or intermittent, narrow-band or
    wideband, and strong or weak.

    Suppose that we have a single interferer impinging onto the telescope 
    array.
    The interfering signal reaches the array with different delays
    $\tau_i$ for each telescope. Assuming processing in narrow sub-bands
    as before, delays translate into phase shifts,\footnote{
For this, the processing bandwidths should be much less than the inverse
of the maximal delay. For example, in WSRT the largest baseline is 3000~m,
corresponding to a maximal delay of $10~\mu$s. Hence the narrow-band
assumption holds for bandwidths less than 100~kHz \cite{leshem99spawc}.}
    and the received signal can be modeled as
$
    x_i(t) = a_i \, s(t) \, e^{-2 \pi \jmath  f \tau_i}
$, or in vector notation,
\[
    \bx(t) = 
    \left[\bea{c} 
	 a_1 e^{-2 \pi\jmath f_c \tau_1} \\ 
	 \vdots \\
	 a_p e^{-2 \pi\jmath f_c \tau_p}
    \ena\right] s(t) \;=:\; \ba s(t) \,.
\]
    Here, $s(t)$ is the baseband signal, and
    $a_i$ represents the telescope gain in the direction of the
    interferer, including any possible attenuation of the channel. 
    Unlike much of the array signal processing literature,
    the $a_i$ are likely to be different for each telescope since the
    interferer is typically in the near field. This implies that it
    impinges on each telescope at a different angle, whereas the response of 
    the telescopes is not omni-directional.
    Hence, the corresponding array response vector $\ba$ is now an
    unknown function.  This vector is also called the {\em spatial
    signature} of the interfering source.  It is slowly time varying,
    and we write $\ba = \ba(t)$.

    Similarly, with $q$ interferers, 
\[
    \bx(t) = \sum_{j=1}^q \ba_j(t) s_j(t)   
    = \mA_s(t) \left[\bea{c} s_1(t) \\ \vdots \\ s_q(t)\ena\right] \,,
    \qquad
    \mA_s(t) = \left[ \ba_1(t) \,, \cdots, \, \ba_q(t)\right] \,.
\]
    The subscript `$s$' is used to distinguish $\mA_s(t)$ from the array
    response matrix of the astronomical sources.

    The corresponding correlation matrix at time $t_k$ is
\[
    \mR_k \;=\; {\rm E} \{ \bx(t_k) \bx(t_k)\rH \}
      \;=\; (\mA_s)_k (\mR_s)_k (\mA_s\rH)_k 
    \,.
\]
    The $q\times q$-matrix $(\mR_{s})_k = \rE\{\bs(t_k)\bs\rH(t_k)\}$
    depends on the correlational properties of the
    interfering signals. For independent interferers, it will be a diagonal
    matrix, with the $q$ interfering powers on the diagonal.

    How well an empirical estimate $\mRh_k$ fits to $\mR_k$ depends on the
    stationarity of the scenario, and is open to discussion.  For various
    reasons (mobile interferers with multipath fading, fixed interferers
    such as TV stations moving through the varying sidelobes of the
    rotating telescopes, fringe corrections of up to 10~Hz), the
    stationarity of $\va_i$ is often limited to about 10-100~ms.
    In the rest of the paper, we make the assumption that indeed the
    available $\mRh_k$ are obtained over stationary periods.
    In summary, the overall model including astronomical signals, 
    array imperfections,  interference and noise is given by:
\begin{equation}
    \mR_k \;=\; \mgG_k \mA_k \mB \mA_k\rH \mgG_k
	\;+\; (\mA_s)_k (\mR_s)_k (\mA_s)_k\rH  
	\;+\; \sigma^2 \mI
    \,,
    \qquad
    k = 0, 1, \cdots
    \,.
\end{equation}
where we assume that the interference term $\mA_s$ is unstructured, and 
$\mbox{rank}(\mA_s)=q<p$.
%
%
%
%
%
%

\section{Spatial filtering}				\label{sec:spatfilt}

   An online interference mitigation system will consist of two stages. As a
   first step the presence of interference is detected. This part is considered
   in \cite{leshem99spawc}, \cite{leshem99ahos} and demonstrated on 
   astronomical data in \cite{leshem99i}.
   In the case of continuous interference it is reasonable to use its  
   spatial signature in order to remove it.
   This leads to {\em spatial filtering} techniques.

\subsection{Projecting out the interferer}

   Let us assume that we have obtained a covariance matrix $\mR$, 
   which contains the rather weak covariance matrix of the astronomical 
   sources (visibilities) $\mR_v=\mgG_k \mA_k \mB \mA_k\rH \mgG_k$, 
   plus white noise.\footnote{In this section 
we consider a single covariance matrix so we drop the index $k$.}
   Suppose also that there is an interferer with power $\sigma^2_s$:
\[
   \mR \;= \; \mR_v \;+\; \sigma^2_s \ba \ba\rH \; + \; \sigma^2 \mI  \,.
\]

   Assuming that $\va$ is known it is possible to null all energy with 
   spatial signature $\ba$. To this end, we can introduce the projection 
   matrices
\[
   \bP_\ba = \ba(\ba\rH \ba)^{-1} \ba\rH
   \,,\qquad
   \bP_\ba^\perp = \mI - \ba(\ba\rH \ba)^{-1} \ba\rH \,.
\]
   It is easily seen that $\bP_\ba^\perp \ba = 0$, so that if we apply
   $\bP_\ba^\perp$ as a spatial filter to $\mR$, we obtain
\begin{equation}				\label{eq:noiseproj}
      \mRt := \bP^\perp_\ba \mR\bP^\perp_\ba
      \;=\; \bP^\perp_\ba \mR_v  \bP^\perp_\ba \;+\; \sigma^2 \bP^\perp_\ba \,.
\end{equation}
   Thus, the interference is completely removed. At the same time, the
   visibility matrix is modified by the projections, and the noise is
   not white anymore, since one dimension is missing.  The imaging
   stage has to be aware of this, which is the topic of section
   \ref{sec:imag}.

   This idea is also applicable to multiple narrowband interferers,
   and we do not need to know the spatial signatures of the interferers
   in advance. Indeed, if the total number of interferers inside a subband
   is less than $q<p$, an eigenvalue decomposition allows to estimate
   the corresponding ``interference subspace'' spanned by the spatial
   signatures from the data covariance matrix, and subsequently project
   out this subspace.

   Thus let $\mR=\mU \mLambda \mU\rH$ be the eigendecomposition of $\mR$. 
   For the purpose of interference cancellation we assume that the sky
   sources are weak: $\mR_v \ll \sigma^2 I$, and thus their influence
   can be ignored in the eigendecomposition.
   Let $\mU = \left[\mU_s \ \mU_n \right]$ where $\mU_s$ is $p \times q$ and 
   contains the eigenvectors corresponding to the $q$ largest eigenvalues,
   and $\mU_n$ collects the remaining eigenvectors.
   In the noise free case, $\mR$ has rank $q$ and
   $\mR = \mA_s \mR_s \mA_s\rH = \mU_s \mgL \mU_s\rH$.
   In the noisy case, $\mR=\mA_s \mR_s \mA_s\rH+\gs^2\mI$ with eigenvalue
   decomposition
\begin{equation}				\label{eq:eignoisy}
\bea{lcl}
    \mR
    &=& \mU_s \mLambda_s \mU_s\rH + 
	\sigma^2 [\mU_s \quad \mU_n] [\mU_s \quad \mU_n]\rH
    \\
    &=& [\mU_s \quad \mU_n] 
	  \left[\bea{c|c} 
	      \mLambda_s + \sigma^2 \mI_q & 0 \\ \hline 0 & \sigma^2 \mI_{p-q}
	  \ena\right]
	  \left[\bea{c} \mU_s\rH \\ \mU_n\rH\ena\right] 
\ena
\end{equation}
   Therefore, the smallest eigenvalue ($\gs^2$) has multiplicity $p-q$, and 
\begin{equation}					\label{eq:noisess}
    \spann(\mU_s) = \spann(\mA_s) \,, 
    \qquad \mU_n\rH \mA_s = 0 \,.
\end{equation}
   We refer to $\mU_s$ as the interference subspace.
   According to (\ref{eq:noisess}), $\mA_s \perp \mU_n$, so that
   $\bP_{\mA_s}^\perp = \mU_n \mU_n\rH$.
   Thus, the eigenvalue decomposition of $\mR$ allows us to detect the
   number of interferers (from the number of repeated small
   eigenvalues) and to identify the projection matrix
   $\bP_{\mA_s}^\perp = \mU_n \mU_n\rH$ to project them out, as in
   (\ref{eq:noiseproj}).  Note that we do not have to know $\mA_s$.
   This hinges upon the fact that the noise covariance is white (in
   general: known), and the visibility matrix $\mR_v$ is insignificant
   at these time scales (otherwise, it might disturb the eigenvalue
   decomposition).


   In practice, we only have a sample estimate $\mRh$ of $\mR$.
   The eigenvalue decomposition of this matrix,
\[
    \mRh = \mUh_s \mgLh_s \mUh_s\rH+\mUh_n \mgLh_n \mUh_n\rH.
\]
   gives a maximum likelihood estimate of $\mU_n$ \cite{anderson58}. 

    One might be worried that if we use the estimated subspace
    for the projections, it might leave correlated residual
    interference components that eventually will show up in the final image. 
    This is in fact not the case, as we will
    now demonstrate that the residual interference is spatially white.

\subsection{Residual interference after projections} \label{sec:spatial_effect}

    A perturbation analysis of the eigenvalue decomposition
    allows to obtain asymptotic expressions for the residual 
    interference in the covariance matrices after spatial filtering. 
    To this end, we utilize the following theorem from
    \cite{sharman84}, a proof of which is given in \cite{stoica89}.

\bthm
\label{th:stoica89}
    Let $\mRh$ be the sample covariance matrix based on $N$ samples of 
    a $p-$dimensional complex Gaussian random process, 
    with zero mean and covariance 
    $\mR$. Let $\mR = \mU\mgL\mU\rH$ be an eigenvalue decomposition of
    $\mR$, with  $\mU = [\vu_1,\ldots,\vu_p]$ unitary, and
    $\mgL = \diag[\gl_1, \cdots, \gl_p]$, 
    where $\gl_1 > \ldots > \gl_q > \sigma^2$,
    and $\gl_{q+1} = \cdots = \gl_p = \sigma^2$.
    Let $\mU_s =\left[ \vu_1,\ldots,\vu_q \right]$, then
    for $m>q$ we have that $\mP_{\mU_s}\vuh_m$ is asymptotically a zero-mean
    Gaussian random process with variance determined by 
\begin{equation} 						\label{res_cov}
    \rE\left\{(\mP_{\mU_s}\vuh_i)(\mP_{\mU_s}\vuh_j)\rH\right\}
    = 
    \frac{\gs^2}{N}
    \left[ 
	\sum_{n=1}^q\frac{\gl_n}{(\gl_n-\gs^2)^2}\vu_n\vu_n\rH
    \right] \gd_{i,j}   \;+\; o(N^{-1}) \,,
    \qquad i,j > q \,.
\end{equation}
\ethm

    For simplicity, let us specialize to the case of $q=1$ narrowband 
    interferer, with power $\sigma_s^2$ per sample and
    spatial signature $\ba$ normalized to $\|\ba\|^2 = p$.  
    In this case, $\lambda_1 = p \sigma_s^2 + \sigma^2$, $\vu_1 =
    \ba/\|\ba\| = \ba/\sqrt{p}$, and (\ref{res_cov}) gives
\[
    \rE\left\{\|\mP_{\vu_1}\vuh_m\|^2\right\}
    \;=\; \frac{\gs^2}{N} 
	  \frac{p \sigma_s^2 + \sigma^2}{(p\sigma_s^2)^2} 
      \,,\qquad m > 1  \,.
\]
    Now note that $\|\mP_{\vu_1}\vuh_m\|^2 = \|\vu_1\vu_1\rH\vuh_m\|^2 =
    \|\vu_1\rH\vuh_m\|^2 = \|\vuh_m \vuh_m\rH \vu_1\|^2 =
    \|\mP_{\vuh_m}\vu_1\|^2$, so that
\[
    \rE\left\{\|\mP_{\vuh_m}\ba\|^2\right\} \;=\; \,
    p\rE\left\{\|\mP_{\vuh_m}\vu_1\|^2\right\} \;=\;
    p\rE\left\{\|\mP_{\vu_1}\vuh_m\|^2\right\}\,.
\]
Similarly we get for $m \not = n$,
$
    \rE\left\{(\mP_{\vuh_m}\vu_1)\rH (\mP_{\vuh_n}\vu_1)\right\} \;=\; 0.
$

    If we define the input Interference to Noise ratio (INR) as
    $\INR = \sigma_s^2/ \sigma^2$, then we finally obtain
\begin{equation} 					\label{eq:res_inta}
    \rE\left\{\|\mP_{\vuh_m} \ba \|^2\right\}
    \;=\; p \frac{\gs^2}{N} 
	  \frac{p \sigma_s^2 + \sigma^2}{(p\sigma_s^2)^2}  
    \;=\; \frac{p}{N} 
	  \frac{p\, \INR + 1}{(p\,\INR)^2}  
      \,,\qquad m > 1 
\end{equation}
\begin{equation} 					\label{eq:res_intb}
\Rightarrow\qquad
    \rE\left\{\|\mP_{\bah}^\perp \ba \|^2\right\}
    \;=\; \rE\left\{\|\mP_{\vuh_1}^\perp \ba \|^2\right\}
    \;=\; \frac{p(p-1)}{N} 
	  \frac{p\, \INR + 1}{(p\,\INR)^2}  
    \,.
\end{equation}
    Let us assume that the estimate $\bah_1$ is approximately independent
    of the interfering signal.
    The residual INR at the output after spatial filtering is then
\begin{equation}					\label{residual_power}
    \INR' := \frac{\gs_s^2}{\gs^2}
	\frac{\rE\left\{\|\mP_{\bah_1}^\perp \ba \|^2\right\}}{p-1}
	\;=\; \frac{1}{N} \left( 1 + \frac{1}{p\, \INR}\right)
\end{equation}

    These expressions are very satisfactory. Indeed note from
    (\ref{eq:res_inta}) that the expected residual interference
    power is the same in each of the directions $\vuh_m$, 
    which together form an orthonormal basis of the projected space. 
    This means that the residual interference is
    spatially white within the projected space (up to second order 
    (in $\frac{1}{N}$) effects),
    and only increases the effective noise power without adding spatial
    features to it.  The effective noise power at the output is
\[
    (\sigma')^2 = \sigma^2 
	    \left\{ 
		1 + \frac{1}{N} \left( 1 + \frac{1}{p\, \INR}\right)
	    \right\} \,.
\]

\begin{figure}
    \begin{center}
	\mbox{\psfig{figure=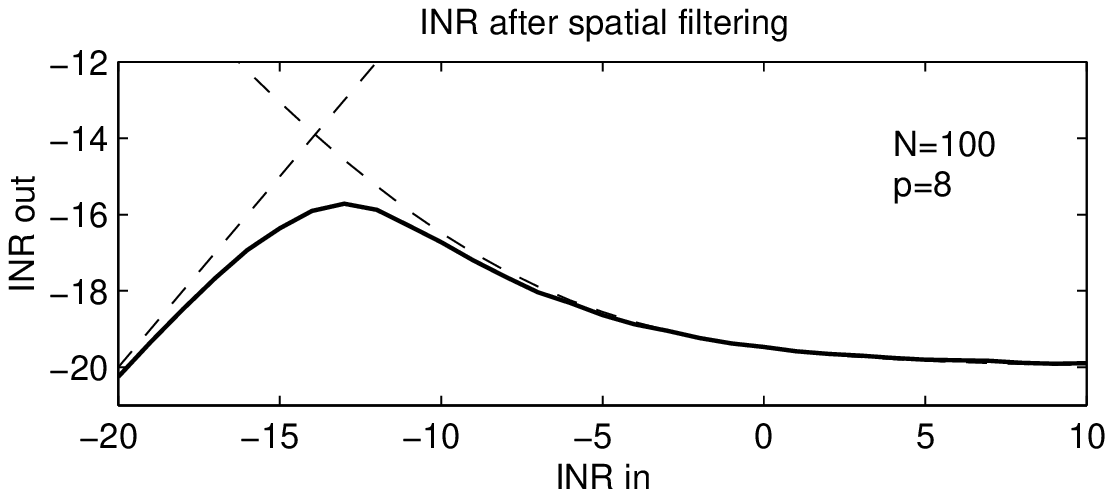,width=0.5\textwidth}}
    \end{center}
    \caption{Residual INR after spatial filtering.}
  \label{Fig:spatfiltinr}
\end{figure}

    Figure \ref{Fig:spatfiltinr} shows the residual interference in a
    simulation, for $N=100$ samples and $p=8$ antennas.
    The reference lines are given by the predicted value in
    equation (\ref{residual_power}),  and the line $\INR' \equiv \INR$.
    Although the predicted value fits very well for sufficiently large
    INR, it is seen that for small INR, (\ref{residual_power}) loses its
    validity. This is because theorem \ref{th:stoica89} is valid only for
    eigenvalues sufficiently above the noise power. For small INRs, the
    estimated interference subspace will be a random vector, and the
    projection will have no effect on the INR.  The cross-over
    point is approximately given by $\INR = \frac{1}{N} +
    \frac{1}{\sqrt{Np}}$.
    The generalization for higher dimensional interference subspace 
    is straightforward using the orthogonality of the interference 
    eigenvectors. Note however that in this case the eigenvectors lose their
    natural interpretation as spatial signatures of the various interferers.



\subsection{Other spatial filtering possibilities}

    Without going into too much detail, we mention a few other
    possibilities for spatial filtering and interference cancellation.
    Suppose there is a single interferer,
\[
    \mR = \mR_v + \sigma_s^2 \ba\ba\rH + \sigma^2\mI \,.
\]

\begin{dashlist}
\item
   {\em Subtraction.}  With an estimate of $\ba$ and an estimate of its
   power, we can try to subtract it from the covariance data:
\begin{equation}				\label{eq:intsubtract}
      \mRt = \mR - \hat{\sigma}_s^2 \bah\bah\rH  \,.
\end{equation}
   Without other knowledge, the best estimate of $\ba$ is the dominant
   eigenvector, $\bu_1$, of $\mR$, and likewise
   the best estimate of $\sigma_s^2$ is $\lambda_1 - \sigma^2$. 
   Since both of these are derived from $\mR$, it turns out to be not too
   different from the projection scheme. Indeed, if we look at
\[
    (\mI - \alpha \bu_1\bu_1\rH) \mR (\mI - \alpha \bu_1\bu_1\rH)
    \;=\; \mR - \bu_1 \bu_1\rH \lambda_1(2 \alpha - \alpha^2)
\]
   we can make it equal to (\ref{eq:intsubtract}) by
   selecting $\alpha$ such that $\lambda_1(2 \alpha - \alpha^2) =
   \hat{\sigma}_s^2$.   The projection scheme had $\alpha = 1$.
\item
   {\em Spatial whitening.}  In this scheme, we try to 
   make the interference plus noise white again.  This component is
   equal to $\sigma_s^2 \ba\ba\rH + \sigma^2\mI$, so we pre- and
   post-multiply with square-root factors of it:
\[
\bea{lcl}
   \mRt &=& (\hat{\sigma}_s^2 \bah\bah\rH + \hat{\sigma}^2 \mI)^{-1/2}
         \mR  (\hat{\sigma}_s^2 \bah\bah\rH + \hat{\sigma}^2 \mI)^{-1/2}
\\
         &=& (\cdot)^{-1/2} \mR_v (\cdot)^{-1/2} + \mI
	 \,.
\ena
\]
\item
    {\em Subtraction of a reference signal.}
    If we have a reference antenna that receives a `clean' copy of the
    interfering signal, then we might try to subtract this reference signal
    from the telescope signals.  There are many adaptive schemes for doing
    so, \eg LMS or RLS. This solution involves a separate receiver for each 
    interferer. 
    To shorten the filter lengths sub-band processing is recommended,
    as otherwise the adaptation might be slow for wideband signals.

    This filtering scheme is similar to the first-mentioned 
    subtraction scheme, except that the spatial signature $\bah$ of the
    interferer is computed from correlations with the reference antenna.
\end{dashlist}

    Note that each of these filtering schemes can be described as a
    linear operation on the entries of the observed data covariance
    matrix $\mRh_k$.  In future equations, we will denote the linear
    operation by $\mL_k$, and the output after filtering $\mRt_k= \mL_k
    \mR_k \mL_k\rH$.


\section{Fourier imaging after spatial filtering}	\label{sec:imag}

    In the previous sections, we discussed spatial filtering techniques.
    It was shown that an attractive scheme for removing the
    interference is by projecting it out. However, by doing so we
    replace the observed visibilities $V(u_i,v_i)$ in the matrix
    $\mR_v$ by some (known) linear combination.  In this section, we
    discuss the implications of this for the imaging.

\subsection{Classical inverse Fourier imaging} 		\label{sec:fourierimag}

    The relation between sky brightness $I(\ell,m)$ and
    visibilities $V(u,v)$ (where $u$, $v$ are taken at frequency $f$) is
\[
    V(u,v) = \iint\; I(\ell,m)\; e^{- 2\pi \jmath (u\ell + v m)}\; d\ell\; d m
\]
    We have measured $V$ on a discrete set of baselines $\{(u_i,v_i)\}$.
    The ``dirty image'' (a lumpy image obtained
    via direct Fourier inversion possibly modified
    with some weights $c_i$) is defined by
\begin{equation}					\label{eq:dirtyimage}
    I_D(\ell,m) \;:=\; \sum_i \;c_i\; V(u_i,v_i) \;e^{2\pi \jmath (u_i \ell + v_i m)}
\end{equation}
    It is equal to the 2D convolution of the true image $I$ with a
    point spread function known as the ``dirty beam'':
\[
  \bea{lcl}
      I_D(\ell,m) &=& \sum_i \; c_i \; V(u_i,v_i) \; e^{2\pi \jmath(u_i \ell + v_i m)}
      \\
      &=& \sum_i \; c_i 
	  \left[ 
	      \iint \; I(\ell',m') \; e^{-2\pi \jmath (u_i \ell' + v_i m')} 
	      \; d\ell' \; dm' 
	  \right] 
	  e^{2\pi \jmath(u_i \ell + v_i m)}
      \\
      &=& \iint \; I(\ell',m') 
	  \left[ 
	      \sum_i \; c_i \; e^{2\pi \jmath (u_i (\ell-\ell') + v_i (m-m'))} 
	  \right] 
	  d\ell' \; dm' 
      \\
      &=& \iint \; I(\ell',m')\;   B_0(\ell-\ell', m-m')\;  d\ell'\; dm'\\
      \ena
\]
    or
\[
    I_D \;=\; I \ast B_0\,,
    \qquad
    B_0(\ell,m) \; :=\;  \sum_i \; c_i\;  e^{2\pi\jmath (u_i \ell + v_i m)} 
\]
    $B_0$ is the dirty beam, centered at the origin. The weights
    $\{c_i\}$ are arbitrary coefficients designed to obtain an
    acceptable beam-shape, with low side lobes, in spite of the
    irregular sampling.

    Specializing to a point source model, 
    $I(\ell,m) = \sum_l \,I_l \,\delta(\ell-\ell_l,m-m_l) $
    where $I_l$ is the intensity of the source at location $(\ell_l,m_l)$,
    gives
\[
    V(u,v) \;=\; \sum_l\; I_l\; e^{- 2\pi \jmath(u\ell_l + v m_l)}
\]
\[
    I_D(\ell,m) \; =\;  \sum_l \; I_l\;  B_0(\ell-\ell_l,m-m_l)
\]
    Thus, every point source excites the dirty beam centered at its location
    $(\ell_l,m_l)$.

    From the dirty image $I_D$ and the known dirty beam $B_0$, the desired
    image $I$ is obtained via a deconvolution process.  A popular method
    for doing this is the CLEAN algorithm \cite{hogbom74}. 
    The algorithm assumes that $B_0$ 
    has its peak at the origin, and consists of a loop in which a
    candidate location $(\ell_l,m_l)$ is selected as the largest peak 
    in $I_D$, and subsequently a small multiple of $B_0(\ell-\ell_l,m-m_l)$
    is subtracted from $I_D$. The objective is to minimize the residual,
    until it converges to the noise level. A short description of the 
    algorithm is given in table \ref{tab:CLEAN}. The parameter
    $\grg \le 1$ is called the 
    loop gain and serves the purpose of interpolation over the grid,
    $\lambda_l$ is the estimated power of the source.

\subsection{Inverse Fourier imaging after projections}\label{sec:invfourproj}

    If we take projections or any other linear combination 
    $[c_{ij}]$ of the
    visibilities $\{V(u_i,v_i)\}$ during measurements, as in section
    \ref{sec:spatfilt}, we have instead available
\[
    Z(u_i,v_i) \; = \; \sum_j \; c_{ij} \; V(u_j,v_j)
\]
    The coefficients $c_{ij}$ are connected to the linear operations
    $\{\mL_k\}$ of section \ref{sec:spatfilt}, and $Z(u_i,v_i)$ are the
    samples contained in the collection $\{\mRt_k\}$.

    Suppose we compute the dirty image in the same way as before, but now
    from $Z$,
\[
    \bea{lcl}
	I_D(\ell,m) &:=& \sum_i \;  Z(u_i,v_i) \; e^{2\pi \jmath (u_i \ell + v_i m)} \\
	&=& \sum_i\sum_j \; c_{ij} \;  V(u_j,v_j) \; e^{2\pi \jmath(u_i \ell + v_i m)}
    \,.
    \ena
\]
    Then
\[
    \bea{lcl}
	I_D(\ell,m) 
	&=& \sum_i\sum_j \; c_{ij} \;  V(u_j,v_j)\;  e^{2\pi \jmath(u_i \ell + v_i m)}
      \\
      &=& \sum_i \sum_j \; c_{ij}
	  \left[ 
	      \iint \; I(\ell',m') \; e^{-2\pi \jmath(u_j \ell' + v_j m')}\;  d\ell' \; dm' 
	  \right] 
	  e^{j(u_i \ell + v_i m)}
      \\
      &=& \iint \; I(\ell',m') 
	  \left[ 
	      \sum_i\sum_j \; c_{ij} \; e^{-2\pi \jmath(u_j \ell' + v_j m')} 
			  \;  e^{2\pi \jmath(u_i \ell + v_i m)}
	  \right] 
	  d\ell' \; dm' 
      \\
      &=& \iint \; I(\ell',m') \;  B(\ell,m,\ell',m') \; d\ell'\; dm'
   \ena
\]
    where
\[
    B(\ell,m,\ell',m') \; :=\; 
	\sum_i\sum_j \; c_{ij} \; e^{-2\pi \jmath(u_j \ell' + v_j m')}
			       \; e^{2\pi \jmath(u_i \ell + v_i m)}
    \,.
\]
    Thus, the dirty image is again obtained via a convolution, but 
    the dirty beam is now space-varying.  $B(\ell,m,\ell',m')$
    is a beam centered at $(\ell',m')$ and measured at $(\ell,m)$.

    With a point source model,
\[
    I_D(\ell,m) \; =\;   \sum_l \; I_l \; B(\ell,m,\ell_l,m_l)
    \; = \; \sum_l \; I_l \; B_l(\ell,m)
\]
    where   
\[
    B_l(\ell,m) \; := \; 
	\sum_i\sum_j \; c_{ij} \; e^{-2\pi \jmath(u_j \ell_l + v_j m_l)}
			       \; e^{2\pi \jmath(u_i \ell + v_i m)}
    \,.
\]
    Again, every point source excites a beam centered at its location
    $(\ell_l,m_l)$, but the beams may all be different: they are {\em space
    varying}. Nonetheless, they are completely known if we know the linear
    combinations that we took during observations.  Thus, the CLEAN
    algorithm in table \ref{tab:CLEAN} can readily be modified to take the
    varying beam shapes into account: simply replace $B_0(\ell,m)$ by
    $B_l(\ell,m)$ everywhere in the algorithm.  Some remaining issues are
\begin{enumerate}
\item It is not a priori guaranteed that the main peak of $B_l(\ell,m)$
     is indeed centered at $(\ell_l,m_l)$.
\item The noise is not necessarily white, and the coloring should be taken
    into account.
\item The computational complexity is increased since we have to construct
     $B_l(\ell,m)$ for every point $(\ell_l,m_l)$.
\end{enumerate}
    The first two points are addressed in section \ref{sec:cleanspatfilt}.

\begin{figure}
\begin{center}
    \mbox{\psfig{figure=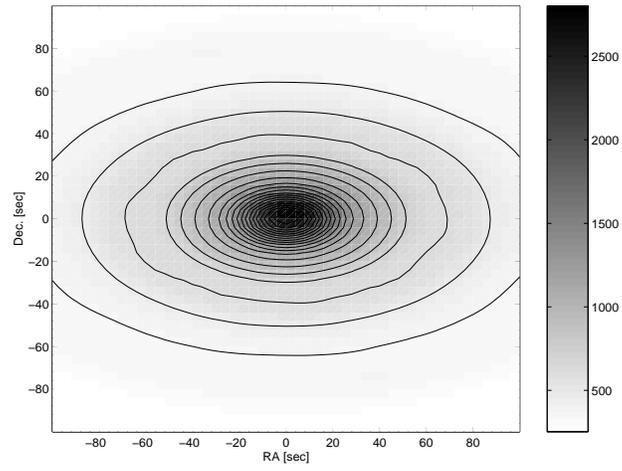,width=0.45\textwidth}}
\end{center}
    \caption{Classical dirty beam $B_0(\ell,m)$, no interference suppression.} 
    \label{clean_beam}
\end{figure}

\begin{figure}
\begin{center}
    \mbox{\psfig{figure=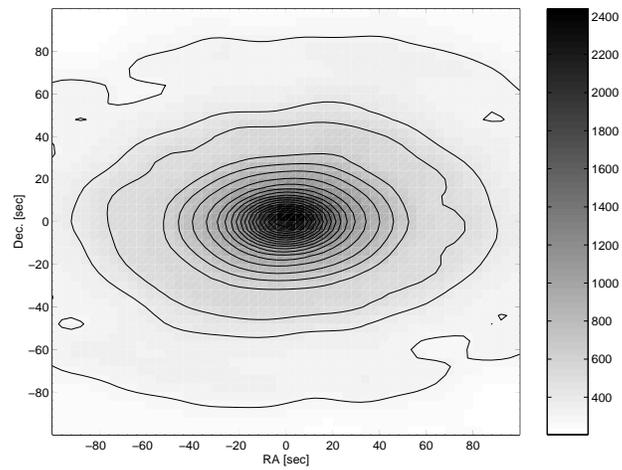,width=0.45\textwidth}}
\end{center}
    \caption{Dirty beam resulting after spatial filtering.
     Beam $B(\ell,m,0,0)$ for a source at the center of the field of view.} 
    \label{filter_beam}
\end{figure}

\begin{figure}
\begin{center}
    \mbox{\psfig{figure=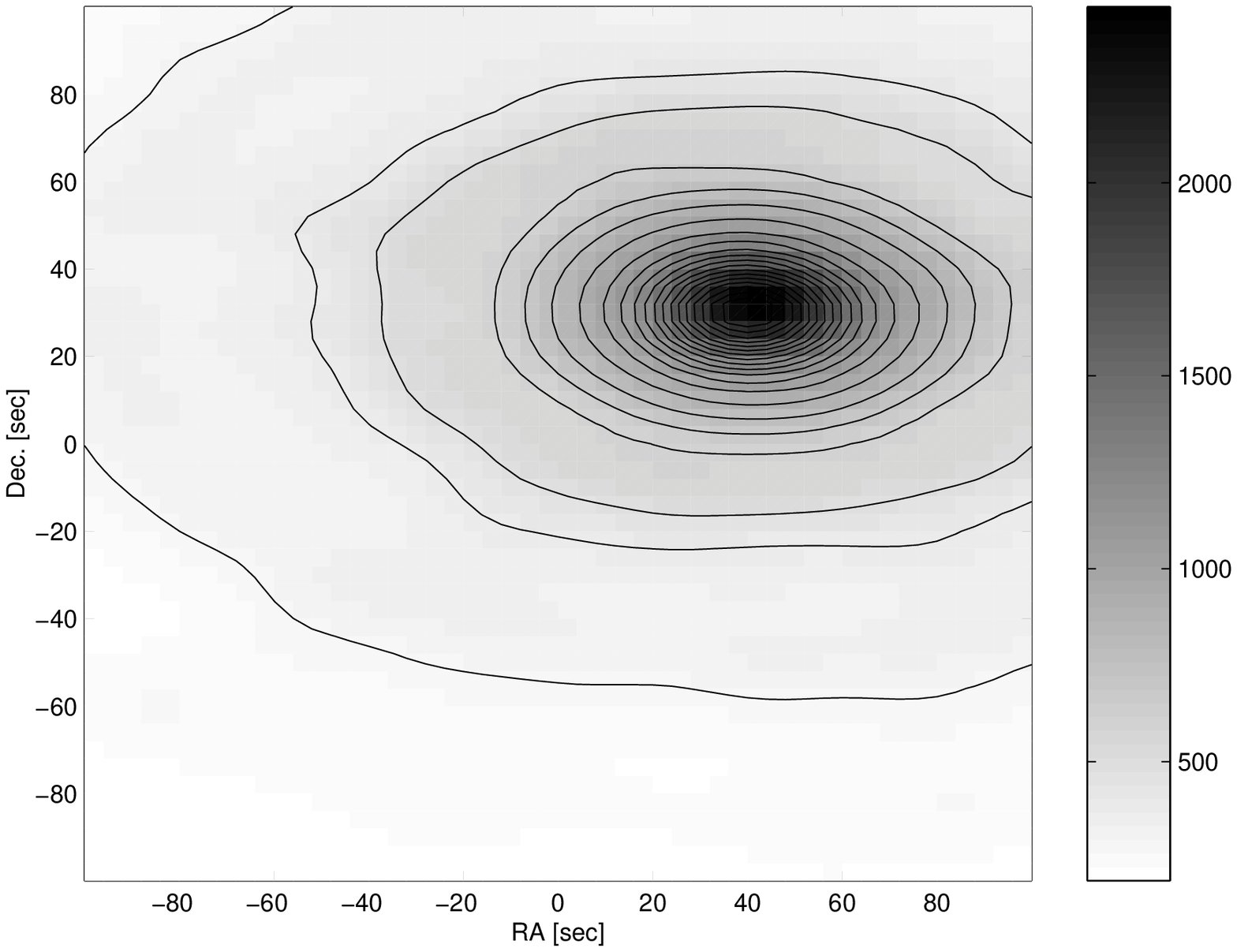,width=0.45\textwidth}}
\end{center}
    \caption{Dirty beam resulting after spatial filtering.  
	 Beam $B(\ell,m,40'',30'')$ for a source at $(40'',30'')$.} 
    \label{filter_beam_side}
\end{figure}

    To demonstrate the spatial variation effect of the dirty beam, we
    have generated an unfiltered dirty beam $B_0(\ell,m)$ (figure
    \ref{clean_beam}), and the beams $B(\ell,m,\ell',m')$ that would
    result after projecting out an interferer with a fixed terrestrial
    location.  We show the latter for a source located at the center of
    the field, \ie $B(\ell,m,0,0)$ (figure \ref{filter_beam}), and for
    a source at $(40'',30'')$, i.e., $B(\ell,m,40'',30'')$ (figure
    \ref{filter_beam_side}).\footnote{
The beams have been computed for $(u_i,v_i)$ samples corresponding to a
WSRT antenna configuration with $p=14$ telescopes and $K=100$
covariance epochs over 12 hours.}
    Note that the spatial projections have modified the shape of the
    beam, in particular the sidelobes, and that the response is not
    constant but varies with the location of the source.   Although the
    changes do not look very dramatic, the differences are in fact
    important: accurate knowledge of the beam shapes
    is essential in the deconvolution step, especially if 
    weak sources are to be detected among sources that are orders of
    magnitude stronger.


\section{Imaging via beamforming techniques}         
\label{sec:bfimag}

    In this section, we reformulate the classical inverse-Fourier
    imaging technique and the CLEAN algorithm for deconvolution in
    terms of a more general iterative beamforming procedure.  This is
    possible since we have a parametric point-source model, and the
    prime objective of the deconvolution step is to estimate the
    location of the point sources.  The interpretation of the
    deconvolution problem as one of direction-of-arrival (DOA)
    estimation allows access to potentially a large number of
    algorithms that have been developed for this application.

\subsection{CLEAN and sequential beamforming} 

    We set out by showing how CLEAN can be interpreted as a 
    sequential beam-forming procedure.

    Let us assume that we have available a collection of measured
    covariance matrices $\mRh_k$, obtained at times $t_k$ with $k=1,
    \cdots, K$, and let us assume the parametric model of
    (\ref{self_cal_model}), \ie
\[    
    \mR_k = \mA_k \mB \mA_k\rH + \sigma^2 \mI \,.
\]
    Here, the unknown parameters
    are the source locations $\vs_l = (\ell_l, m_l)$, $l = 1, \cdots, d$
    in each of the $\mA_k$, and the source brightness $I_l$ in $\mB$. 
    A natural formulation for the estimation of these parameters
    is to pose it as the solution of a LS cost function, given by
\begin{equation} 					\label{CLEAN_LS}
    [\{\hat{\vs_l}\} ,\mBh] 
    =
    \argmin_{\{\vs_l\},\mB}
    \sum_{k=1}^K\;
    \|\; \mRh_{k} - \mA_{k}(\{\vs_l\})\, \mB \mA\rH_{k}(\{\vs_l\})\, 
    - \gs^2 \mI \;\|_F
\end{equation}
    ($\mB$ is constrained to be diagonal with positive entries.)
    This is recognized as the same model as used for DOA estimation in array
    processing.  Note however that the array is moving ($\mA_{k}$ is
    time-dependent), and that there are many more sources than the
    dimension of each covariance matrix.

    In this notation, the image formation in section \ref{sec:fourierimag} 
    can be formulated as follows.  Recall from (\ref{eq:visdef}) 
    and (\ref{eq:vadef}) that
\[
    V(u_{ij}(t_k), v_{ij}(t_k)) \;\equiv\; r_{ij}(t_k)\; \equiv\; (\mR_k)_{ij}
    \,,\qquad
    k = 1, \cdots, K
\]
\[
    \va_k(\ell,m)
    =
    \left[ \bea{c}
	e^{-2\pi \jmath(u_{10}(t_k) \ell + v_{10}(t_k) m)} \\
	\vdots \\
	e^{-2\pi \jmath(u_{p0}(t_k) \ell + v_{p0}(t_k) m)}
    \ena \right]
\,,\qquad
\bea{lcl}
    u_{ij} &=& u_{i0} - u_{j0}\,,\\
    v_{ij} &=& v_{i0} - v_{j0}
\ena
\]
    Thus, if we write $I_D(\bs) \equiv I_D(\ell,m)$ and 
    $\ba_k(\bs) \equiv \ba_k(\ell,m)$,
    we can rewrite the dirty image (\ref{eq:dirtyimage}) as
\[
\bea{lcl}
    I_D(\bs) &=& \sum_{i,j,k} V(u_{ij}(t_k), v_{ij}(t_k)) 
	    e^{2\pi \jmath(u_{i0}(t_k) \ell + v_{i0}(t_k) m)} 
	    e^{-2\pi \jmath(u_{j0}(t_k) \ell + v_{j0}(t_k) m)} 
\\ 
    &=& \sum_{i,j,k}   \; (\mR_k)_{ij} (\bar{\ba}_k(\bs))_i (\ba_k(\bs))_j 
\\
    &=& \sum_k \; \ba_k\rH(\bs) \mR_k \ba_k(\bs) \,.
\ena
\]
    (We omitted the optional weighting. Also note that, with noise, 
    we have to replace $\mR_k$ by $\mR_k-\sigma^2 \mI$.)
    The iterative beam removing in CLEAN can now be posed as an
    iterative LS fitting between the sky model and the observed visibility
    \cite{schwarz78}. 
    Finding the brightest point $\bs_0$
    in the image is equivalent to trying to find a 
    point source using classical Fourier beamforming, \ie,
\begin{equation}
\label{CLEANbf}
    \vsh_0 = \argmax_{\vs} \sum_{k=1}^K 
       \va_k\rH(\vs) \left(\mR_k - \gs^2\mI\right) \va_k(\vs) \,.
\end{equation}
    Thus, the CLEAN algorithm can be regarded as a generalized classical
    sequential beamformer, where the brightest points are found one by one,
    and subsequently removed from $\mR_k$
    until the LS cost function (\ref{CLEAN_LS}) is minimized.
    An immediate consequence is that the estimated source locations
    will be biased: a well known fact in array processing.  When the
    sources are well separated the bias is negligible compared to the
    standard deviation, otherwise it might be significant.  This gives
    an explanation for the poor performance of the CLEAN in imaging
    extended structures (see e.g., \cite{perley89}). Another enhancement in 
    the CLEAN can be made by turning it into an iterative scheme rather than 
    sequential. In this case after estimating all point sources, we put one
    source at a time back into the data, and re-estimate it, using the 
    estimates of the other point sources. This will improve the LS fit, and 
    can easily be proved to converge. This approach is similar to the alternating projections approach for computing the deterministic ML DOA estimator 
\cite{WZ}.

\subsection{Minimum variance beamforming approaches}

    Once we view image formation/deconvolution as equivalent to 
    direction-of-arrival (DOA) estimation with a
    moving array, we can try to adapt various other DOA estimators for
    handling the image formation.  In particular the deflation approach
    used in the CLEAN algorithm can be replaced by other source
    parameters estimators. One approach that seems particularly relevant in
    this context is the Minimum-Variance Distortionless Response (MVDR)
    method of beamforming \cite{capon69}.  The major new aspect here is the
    fact that the array is moving and that there are more sources than
    sensors.

    Instead of working with the dirty image 
    $I_D(\bs) = \sum_k \; \ba_k\rH(\bs) \mR_k \ba_k(\bs)$, 
    the basis for high-resolution beamforming techniques is
    to look at more general ``pseudo-spectra''
\begin{equation} 					\label{eq:MVDRimage}
    I_D'(\bs) \;:=\; \sum_k \; \bw_k\rH(\bs) \mR_k \bw_k(\bs)
\end{equation}
    Here, $\bw_k(\bs)$ is the beamformer pointing towards direction $\bs$,
    and $I_D'(\bs)$ is the output energy of that beamformer.  Previously we
    had $\bw_k(\bs) = \ba_k(\bs)$; the objective is to construct
    beamformers that provide better separation of close sources.

\begin{figure}
\begin{center}
    \mbox{\psfig{figure=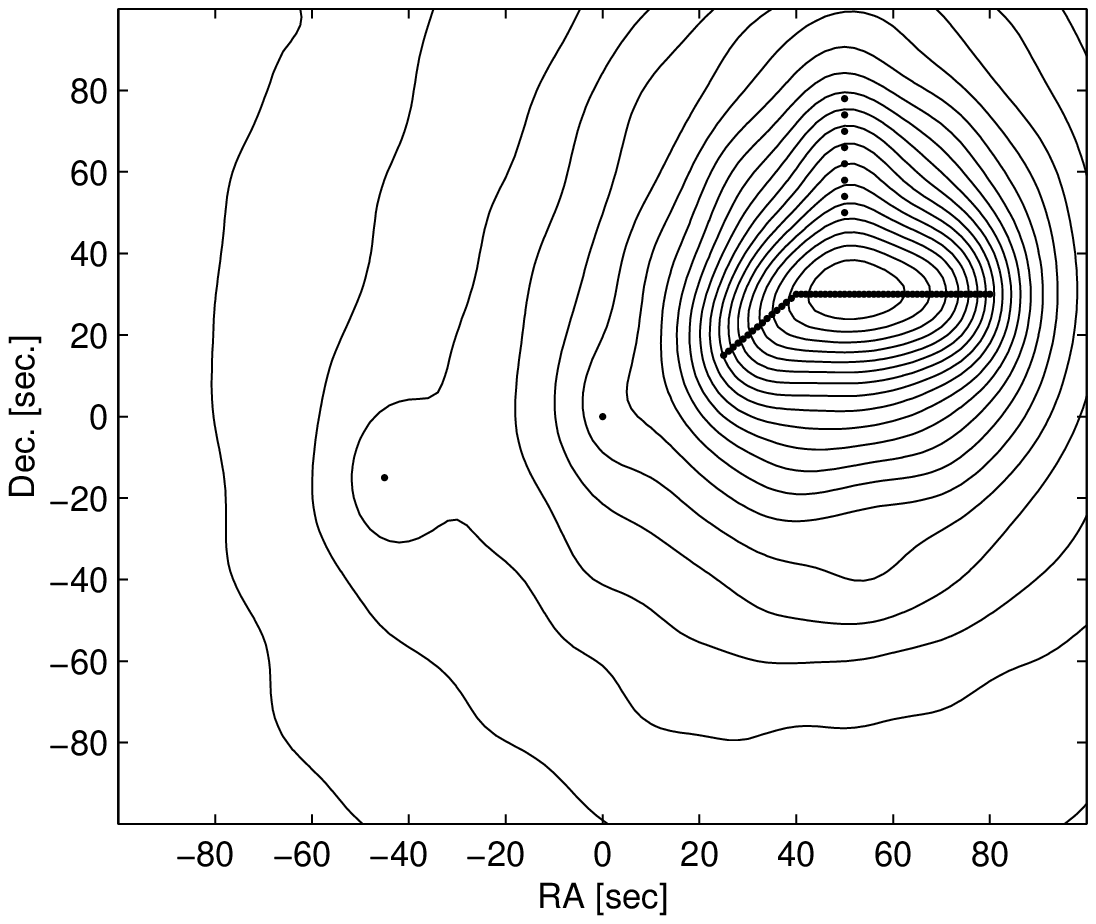,width=0.6\textwidth}}
    \mbox{\psfig{figure=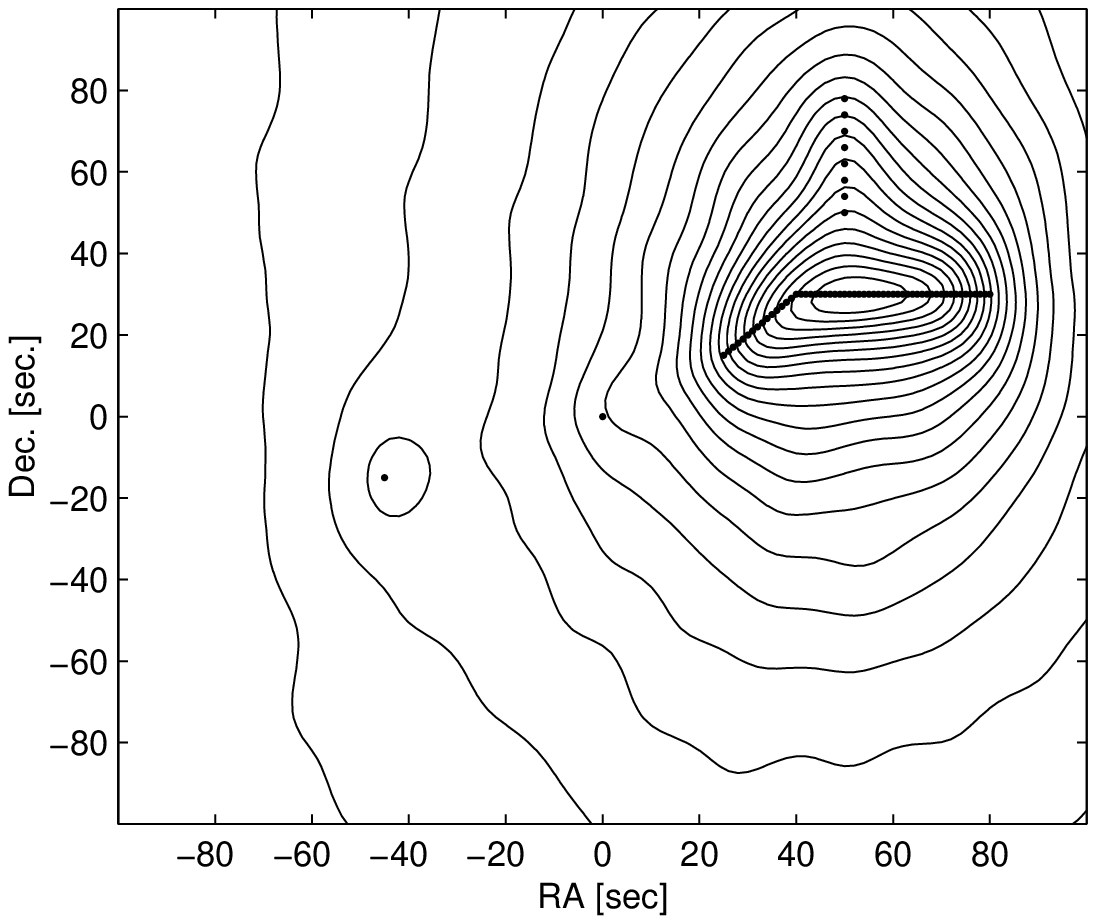,width=0.6\textwidth}}
\end{center}
    \caption{$(a)$ Conventional dirty image; $(b)$ dirty image using MVDR
	beamforming. The dots represented the locations of the point sources modeled.}
    \label{Fig:MVDRimage}
\end{figure}

    A generalized MVDR follows by defining the problem as
    follows:  At each time instance $k$ we would like to generate 
    a weight vector $\vw_k$ which minimizes the output power
    at time $k$ subject to the constraint that we have a fixed response
    towards the look direction $\vs$ of the array, i.e.,
\[
    \vwh_k(\vs) \;=\; \argmin_{\vw_k}\; \vw_k\rH \mRh_k \vw_k 
    \qquad\hbox{\ such that \ } \quad
    \vw_k\rH\va_k(\vs) = 1 \,.
\]
    The solution to this problem is
\begin{equation}
    \vwh_k = \gb_k \mRh_k^{-1} \va_k(\vs)
    \,,
    \qquad\mbox{where } \quad
	\gb_k = \frac{1}{\va_k\rH(\vs) \mRh_k^{-1} \va_k(\vs)} \,.
\end{equation}
    Inserting in (\ref{eq:MVDRimage}) shows that
    the overall spectral estimator is given by
\begin{equation}					\label{eq:MVDRaimage}
    I_D'(\bs) \;=\; 
       \sum_{k=1}^K \;\frac{1}{\va_k\rH(\vs) \mRh_k^{-1} \va_k(\vs)}
    \,.
\end{equation}
    and the locations of the strongest sources are given by the maxima of
    $I_D'(\bs)$.
    It is known that the MVDR has improved resolution compared
    to the classical beamformer which is the basis for the CLEAN algorithm.
    Figure \ref{Fig:MVDRimage} illustrates this by comparing
    a dirty image produced in the classical way to the dirty image
    corresponding to (\ref{eq:MVDRaimage}).  In this simulation,
    we generated an extended structure by placing many point sources
    close to each other.  The MVDR-based imaging produces a much
    sharper result.

    Compared to the CLEAN there is a slight computational loss.
    Another drawback of the MVDR is the fact that the noise
    distribution at the output of the beamformer 
    is not identical towards all directions. Here we can
    use a modification \cite{borgiotti79} which demands that $\vw_k$
    has a functional form form $\vw_k = \gb_k \mRh_k^{-1} \va_k(\vs)$, 
    for some $\gb_k$, but now under the constraint that $\|\vw_k\|^2=1$. 
    This leads to the following estimator:
\[
    I_D''(\bs) \;=\; \sum_{k=1}^K \frac{\va_k\rH(\vs) \mRh_k^{-1} \va_k(\vs)}
			  {\va_k\rH(\vs) \mRh_k^{-2} \va_k(\vs)} 
    \,,\qquad
    \vsh_0 = \argmax_{\vs} \; I_D''(\bs) \,.
\]
    Many other techniques exist for estimating the point source
    locations. A good overview of the various possibilities can be found
    in \cite{krim97}.

\subsection{CLEAN with spatial filtering} 	\label{sec:cleanspatfilt}

    Let us assume now that we have spatially filtered the covariance
    matrices $\mRh_k$ by linear operations $\mL_k$, for example projections.
    If we assume that all the 
    interference is removed by the filtering, the measurement equation becomes
\begin{equation}
    \mRt_{k} \; := \; \mL_k \mR_k \mL_k\rH \;=\;
	\mL_k \left[
	    \mA_{k}(\{\vs_l\}) \mB \mA\rH_{k}(\{\vs_l\}) + \sigma^2 \mI 
	\right]\mL_k\rH \,.
\end{equation}
    This modifies the least squares optimization problem to
\begin{equation} 					\label{proj_cal}
    \left[\{\vsh_l\}, \mBh  \right] \;=\;
    \argmin_{\{\vs_l\} , \mB, \{\mgG_k\}}
    \sum_{k=1}^K \;
    \|\, \mL_k 
	\left(
	   \mRh_{k} \,-\,
	   \mA_{k}(\{\vs_l\})\, \mB \mA\rH_{k}(\{\vs_l\})
	    \,-\, \sigma^2 \mI 
	\right) \mL_k\rH
	\,\|_F \,.
\end{equation}
    The cost function is similar to (\ref{self_cal_cost}) and thus its
    minimization does not pose stronger computational demands.
    Indeed, as we mentioned before in section \ref{sec:invfourproj}, 
    if we follow the classical Fourier-type imaging we end up with a
    deconvolution problem with a space-varying beam, but 
    the CLEAN algorithm is simply extended to take this into account.
    Here, we develop the extension more carefully, taking note of the fact
    that the noise structure after projections is not white anymore.

    In the case of spatially filtered signals the classical beamformer follows
    from the previous by replacing $\va_k(\vs)$ by the effective array
    response $\mL_k \va_k(\vs)$, \ie
\begin{equation} 					\label{clean_proj_fft}
\bea{lcl}
    I_D'(\vs) &=& \sum_{k=1}^K 
	\va_k\rH(\vs) \mL_k\rH 
	\bigl(\mL_k \mRh_k \mL_k  \;-\; \gs^2\mL_k \mL_k\rH\bigr)  
	\mL_k \va_k(\vs)
\\
    &=&
	\sum_{k=1}^K 
	\va_k\rH(\vs) 
	    \left(\mL_k\rH\mL_k\,\mRh_k \mL_k\rH\mL_k \; -\; 
		\gs^2\mL_k\rH\mL_k \mL_k\rH\mL_k \right) 
	\va_k(\vs) 
\\
    &=& \sum_{k=1}^K \va_k\rH(\vs) \mRh'_k \va_k(\vs) \,,
\ena
\end{equation}
where 
\begin{equation}
\label{mRhpdef}
\mRh'_k \;=\; 
	    \mL_k\rH\mL_k\, \mRh_k\, \mL_k\rH\mL_k 
	    \;-\; \gs^2\mL_k\rH\mL_k\, \mL_k\rH\mL_k 
\end{equation}
    Therefore the step of finding the brightest point $\vs_0$ in the
    image can be implemented using FFT in the same way it is implemented
    in the CLEAN algorithm, but acting on $\mRh'_k$ instead of the
    original visibilities.
    Similarly, the contribution of a source at location $\vs_0$ in a single
    covariance matrix $\mRt_k$ 
    is a multiple of $\mL_k \va_k(\vs_0) \va_k\rH(\vs_0)
    \mL_k\rH$, and hence the response in the dirty image $I_D'(\vs)$ 
    is given by 
\begin{equation}
\label{defB}
    B(\bs,\bs_0) :=
	\sum_{k=1}^K 
	    \va_k\rH(\vs) \mL_k\rH 
	    \left( \mL_k \va_k(\vs_0) \va_k\rH(\vs_0)\mL_k\rH  \right)
		\mL_k \va_k(\vs) \,.
\end{equation}
    This is the space-varying beam.   The extended CLEAN algorithm after
    spatial filtering now follows immediately and is given in table
    \ref{tab:clean_filter}.

    To test the algorithm, we have taken an array configuration with
    $p=14$ telescopes as in WSRT, and generated four equal-powered
    point sources centered around right ascension $32^{\circ}$ and
    declination $60^{\circ}$, with a signal to noise ratio of $-20$~dB
    for each of the sources.  To simulate the effect of spatial
    filtering, we placed an interferer at a fixed terrestrial location
    (hence varying compared to the look direction of the array), and
    with $\INR=5$~dB.  $K=100$ sample covariance matrices $\mRh_k$ were
    generated, uniformly spread along 12 hours, and each based on
    $N=1000$ samples.  Figure \ref{Fig:spatialclean}$(a)$--$(c)$ 
    shows the dirty image without interference present, the effect of
    the interferer on the dirty image, and the dirty image after
    estimating and removing the interferer using spatial projections.
    Clearly, with interference present but not removed, the sources are
    completely masked out (note the change in scale between the first
    two figures).  After estimating and projecting out the interferer, 
    in the third image,
    we obtain nominally the same image as in the interference-free case,
    but the sidelobe patterns are different (as we demonstrated before, they
    are in fact space-varying).
    The circles in the third image mark the true location of the point
    sources, and the $+$-symbols mark the locations estimated by the
    extended CLEAN algorithm of table \ref{tab:clean_filter}.  We can
    clearly see that the correct locations have been obtained.  This
    would not be the case with the unmodified CLEAN algorithm.
\subsection{Self-calibration}

    To finish this section, we consider the situation where also the array
    gains $\mgG_k$ are unknown.
    In this case, the model fitting equation without spatial filtering,
    equation (\ref{CLEAN_LS}), generalizes to
\begin{equation} 					\label{self_cal_cost}
    \left[\{\vsh_l\} , \mBh, \{\mgGh_k\} \right]
    =
    \argmin_{\{\vs_l\},\mB,\{\mgG_k\}}
    \sum_{k=1}^K \|\,\mRh_{k} 
	- \mgG_{k} \mA_{k}(\{\vs_l\})\, \mB \mA\rH_{k}(\{\vs_l\})\,\mgG_{k}\rH 
	- \sigma^2 \mI \,\|_F
\end{equation}
    The solution can be obtained by the ``self-cal'' algorithm \cite{pearson84}:
    an alternating least squares algorithm which solves iteratively for the
    parameters $\mB,\,\{\bs_l\}$ by a CLEAN step (with fixed gains), 
    and the gain parameters $\{\mgG_k\}$ by a calibration step (with fixed
    source parameters $\mB, \{\vs_l\}$).

    It has not been noted before in the literature that the latter step admits a
    direct algebraic solution. Indeed, to minimize (\ref{self_cal_cost}) with
    fixed $\{\mA_{k}\}$ and $\mB$, we 
    can minimize separately for each $k$ the related expression
\begin{equation}				\label{self_cal_reducedcost}
    \min_{\{\mgG_k\}} \;
    \|\, \mRh_k - \mgG_k \mA_k \mB \mA\rH_k\mgG_k\rH - \sigma^2 \mI \,\|_F
\end{equation}
    Let $\vg_k$ be the vector of diagonal elements $\gamma_{k,i}$ 
    of $\mgG_k$. 
    Given an estimate of $\mA_k$ and $\mB$ we can define for each $k$ a 
    $(p\times p)$ matrix $\mXh_k$ with entries
\[
    (\mXh_k)_{ij} = 
    \frac{(\mRh_{k}-\sigma^2 \mI)_{ij}}{ (\mA_{k} \mB \mA\rH_{k})_{ij}}
\]
    and fit $\vg_k$ with entries $g_{k,i}$ such that
$
    (\mXh_k)_{ij} = g_{k,i} {\bar g}_{k,j} \,.
$
    In the usual self-calibration algorithm, this equation is solved
    iteratively for all two-by-two sub-matrices of $\mXh_k$ using the
    so-called gain and phase closure relations.  Instead, we note here
    that the problem admits a more elegant solution, since
    in matrix form, we have
\[
    \mXh_k = \vg_k \vg_k\rH \,.
\]
    This asks for the best hermitian rank-one approximation to the matrix 
    $\mX_k$, which is known to be given by $ \vgh_k = \sqrt{\gl_1}\vv_1 $
    where $\gl_1$ is the largest eigenvalue of $\mXh_k$ and $\vv_1$
    its corresponding eigenvector. 

    More work is needed to generalize this to the model equation {\em
    with} spatial filtering.  With fixed $\mA_k$ and $\mB$, the minimization
    problem for the gain parameters is
\begin{equation}				\label{self_cal_spatfilt}
    \min_{\{\mgG_k\}} \;
	\|\, \mL_k\left(\mRh_k \;-\; \mgG_k \mA_k \mB \mA\rH_k\mgG_k\rH 
	\;-\; \sigma^2 \mI\right) \mL_k\rH \,\|_F
\end{equation}
    Let $\vect(\,\cdot\,)$ denote the operation of stacking all columns of
    a matrix into a vector, and let $\otimes$ denote the 
    Kronecker product.
    A general property valid for matrices of compatible size is
    $\vect(\mA\mB\mC) = (\mC^T\otimes \mA)\vect(\mB)$.  Application
    to the present context gives that (\ref{self_cal_spatfilt}) asks for
    the solution in least squares sense of
\[
\bea{lcl}
    (\mLb_k\otimes \mL_k)\,\vect(\mRh_k - \sigma^2 \mI) 
    &=& (\mLb_k\otimes \mL_k) \,\vect(\mgG_k \mA_k \mB \mA\rH_k\mgG_k\rH)
\\
    &=& (\mLb_k\otimes \mL_k)\, \diag(\vect(\mA_k \mB \mA\rH_k)) \,
	(\vgc_k\otimes \vg_k)
\ena
\]
    which has the form
\[
    \vh_k = \mF_k \vx_k \,, \qquad\qquad \vx_k = \vgc_k \otimes \vg_k \,.
\]
    If $\mF_k$ has a left inverse $\mF_k^\dagger$, 
    we can find a unique LS solution for $\vx_k$, \ie $\vxh_k = \mF_k^\dagger
    \vh_k$, and then fit $\vxh_k = \vgc_k\otimes \vg_k$, 
    or equivalently $\mXh_k = \vg_k \vg_k\rH$, where $\mXh_k$ 
    is derived from $\vxh_k$ by unstacking, \ie  $\vect(\mXh_k) = \vxh_k$.
    This is precisely the solution which we obtained before.

    However, in the present case $\mL_k$ is a projection operator and hence
    $\mF_k$ is not invertible.  It is easy to see from examples, such as
    taking $\mL_k = \bigl[ {I \atop 0}{0  \atop 0} \bigr]$, that in these
    cases $\mgG_k$ is not identifiable.
    A solution can be obtained if we make the reasonable assumption
    that $\mgG_k$ is constant over several epochs, say for $k, k+1, \cdots,
    k+M-1$, and that $\mL_k$ is sufficiently varying over this period, for
    example due to multipath or fringe corrections.
    In that case, we obtain
\[
    \left[\bea{@{}l@{}}  \vh_k \\ \vh_{k+1} \\ \;\vdots \\ \vh_{k+M-1} \ena\right]
    = 
    \left[\bea{l}  \mF_k \\ \mF_{k+1} \\ \;\;\vdots \\ \mF_{k+M-1} \ena\right]
    \vx_k\,,
    \qquad
    \qquad
    \vx_k = \vgc_k \otimes \vg_k \,.
\]
    For sufficiently large $M$, the block matrix is tall and of full column
    rank, and has a left inverse.  We can thus solve for an unstructured
    LS estimate $\vxh_k$, and subsequently fit $\vxh_k = \vgc_k\otimes
    \vg_k$.

    The minimal number $M$ of linearly independent matrices $\mL_k$
    which are needed follows from counting
    dimensions.  If we project out $q$ interferers on $p$ antennas, 
    $\mL_k$ has $p-q$ independent rows and $p$ columns.
    Hence $\mF_k$ has $(p-q)^2$ independent rows and $p^2$ columns,
    and we need $M(p-q)^2 \ge p^2$.
    This gives modest requirements: if for $p=14$ we take $M = 2$, then
    we can accept up to $q=4$ interferers; with $M=4$, up to $q = 7$.

    In summary, we have obtained an elegant and computationally
    non-intensive extension of the self-cal algorithm to complement the
    space-filtering CLEAN algorithm.


\section{Maximum likelihood imaging} 				\label{sec:MLE}

\subsection{Maximum likelihood functional} 
    Let us consider the imaging step from a more fundamental viewpoint.
    In principle, the construction of the image using the observed
    correlation matrices and assuming the parametric model
    can be viewed as a parameter estimation problem.
    One of the most important inference methods is the maximum likelihood 
    method: Given a parametric family of probabilistic models for the 
    received data, choose the 
    parameters that maximize the probability of obtaining the observed data.
    This is different than the most probable image approach \cite{frieden72}
    where no parametric model is imposed on the image, leading to 
    maximum entropy image formation.
    Maximum likelihood estimators (MLEs) are known to be
    consistent and asymptotically statistically efficient 
    (i.e., they provide unbiased estimators with minimum variance) 
    under very general conditions, and thus are the
    natural choice for many parameter estimation problems. 

    In deriving the maximum likelihood estimator of the image parameters,
    we need a parametric family of models for the astronomical signals.
    A reasonable assumption regarding the astronomical data is Gaussianity of 
    the temporal samples\footnote{The assumption is valid for continuum  
    emission, while for spectral line observations certain adjustments of the 
    model are needed to include the lines structure.}. This assumption is used
    in current imaging systems which rely only on 
    second order statistics (both temporal and spatial). For simplicity we 
    further assume that the samples are temporally white (valid for the 
    relatively narrow bands processed, while over very large bands the 
    black-body radiation pattern should be taken into account). For further 
    discussion on emission mechanisms, and the resulting physical models of emission the reader is referred to \cite{rohlfs99}. 
    Contrary to the claim in \cite{tan86}, the corresponding MLE 
    is {\em not} equivalent to parametric optimization of the CLEAN
    cost function. 
    Using the discrete point source model we obtain:
\begin{equation}					\label{cov_model}
    \mR_k \;=\; \mgG_k \mA_k(\{\vs_l\})\, \mB \mA_k\rH(\{\vs_l\}) \,\mgG_k\rH 
    \;+\; \gs^2\mI
\end{equation}
    where the $d$ astronomical sources are Gaussian with covariance matrix
    $\mB = \diag[I_1 \,, \cdots\,,I_d]$ and sky coordinates 
    $\{ \vs_l\}_{l=1}^d$,
    and the noise is Gaussian with covariance $\gs^2\mI$.
    Let $\mRh_k$ be the sample covariance matrix during the $k$-th epoch,
    based on $N_k$ collected samples.
    The likelihood of the observations at the $k$-th epoch
    given map parameters $\{\vs_l\},\mB, \sigma^2, \mgG_k$ is then given by 
    \cite{anderson58}:
\begin{equation}
    p(\mRh_k | \{\vs_l\},\mB, \sigma^2, \mgG_k )=\left(\frac{1}{\pi^p|\mR_k|} e^{-\trace(\mR_k^{-1}\mRh_k)}\right)^{N_k}
\end{equation}
    Using all $K$ observation epochs we obtain 
    that the log-likelihood function is 
    given by (after omitting constants) 
\begin{equation} 						\label{eq:lik}
    {\cL}(\mRh_1,\ldots,\mRh_K|\mB,\{\vs_l\},\{\mgG_k\},\gs^2)
    = -\sum_{k=1}^K N_k \log |\mR_k|- \sum_{k=1}^K N_k \trace(\mR_k^{-1}\mRh_k)
\end{equation}
    The MLE is found by
    maximizing (\ref{eq:lik}) over $\mB,\{\vs_l\},\{\mgG_k\},\gs^2$.
    This maximization problem is prohibitively complex and hence some 
    simplifications are needed. In some simplified cases in DOA estimation
    this has been dealt with.
    The Gaussian signals model for a static array with perfect calibration 
    have been considered by \cite{bohme86} which eliminated analytically 
    some of the parameters.
    Derivation of the MLE for a single source in white Gaussian noise for
    the simplified model appeared in \cite{zeira95}. 
    
Since the CLEAN  gives an 
approximate solution to the deconvolution problem we can use it to 
initialize a maximum likelihood search. In this case the CLEAN components 
serves as initial estimates to the MLE and the MLE serves the purpose of fine
focusing of the image, by shifting each point source to its true value. The 
ML  search itself can be done either using a gradient search, a Newton search
based on the Fisher information matrix, or an EM algorithm. In appendix B we 
present the expressions of the gradient and the Hessian needed for the 
optimization. Since a good initialization is very important for optimizing the 
complicated equation (\ref{eq:lik}) we will continue to derive an approximate 
coordinate descent MLE which is computationally simpler, and show its 
connection to CLEAN.

\subsection{Single source in colored noise}

    One simplified approach to solve the MLE (\ref{eq:lik}) which leads to 
    good results in LS problems is the deflation approach or coordinate
    descent algorithm. In this approach the sources are extracted one by one
    and once we have obtained estimates of parameters of all sources  
    we iterate the optimization along each parameter fixing the other 
    parameters. Some examples of this approach are the alternating projections
    algorithm \cite{WZ}, and the estimation of multipath parameters
    \cite{leshem97a}. 

    Our basic approach to remove the effect of previously estimated sources 
    will be to lump their contribution into the ``noise part'' of the 
    covariance matrix. This essentially provides an a-posteriori estimate of
    the noise covariance matrix, before continuing to estimate the next source.
    Hence an important component of the algorithm will be the ML 
    estimation of a single source in colored noise with known covariance 
    matrix.
    In what follows we will derive an approximate ML estimator for this
    specific case. In order to 
    reduce the notational complexity we will restrict ourselves to perfectly 
    calibrated arrays, i.e., $\mgG_k=\mI$ for all $k$, and note that 
    estimating the calibration parameters will be done in a separate stage as 
    it is currently done in the self calibration scheme.
     
    In order to reduce computational complexity even further, we would like to 
    perform the source power estimation (conditioned on the location 
    parameter) analytically. This will result in a 
    two-dimensional search over the field of view for the location  
    with highest likelihood for a point source, a task of moderate complexity. 

Assume that we are given a set of covariance matrices 
$\left\{\mRh_k : k=1,\ldots,K \right\}$ which are sample estimates of 
$\mR_k$ where
\begin{equation}
\label{cov_def}
\mR_k=\gl\va_k(\vs)\va_k\rH(\vs)+\mC_k
\end{equation}
and $\mC_k$ is the noise covariance matrix (assumed to be known).

The log-likelihood function is given by:
\begin{equation}
\label{loglik_single}
{\cL}(\mRh_1,\ldots,\mRh_K|\gl,\vs)=-\sum_{k=1}^K N_k \left[ 
\log|\mR_k| +\tr\left(\mR_k^{-1}\mRh_k\right) 
\right].
\end{equation}
Substituting (\ref{cov_def}) into (\ref{loglik_single}) we obtain
\[
{\cL}(\mRh_1,\ldots,\mRh_K|\gl,\vs)=-\sum_{k=1}^K N_k \left[ 
\log|\gl\va_k(\vs)\va_k\rH(\vs)+\mC_k|+\tr\left(\left(\gl\va_k(\vs)
\va_k\rH(\vs)+\mC_k\right)^{-1}\mRh_k\right)
\right].
\]
Further manipulation  using $\tr(\mA\mB)=\tr(\mB \mA)$ and 
$|\mA\mB|=|\mA| |\mB|$ yields
\[
\begin{array}{lll}
{\cL}(\mRh_1,\ldots,\mRh_K|\gl,\vs)= & -\sum_{k=1}^K N_k & 
\Bigl[\log|\gl\mC^{-1}\va_k(\vs)\va_k\rH(\vs)+\mI|
\\
& & \left.
+\log|\mC_k|+\tr\left(\left(\gl\mC_k^{-1}\va_k(\vs)\va_k\rH(\vs)+\mI
\right)^{-1} \mC_k^{-1}\mRh_k\right)
\right].
\end{array}
\]
Since $\log|\mC_k|$ does not depend on the parameters, the 
maximum likelihood estimate is given by minimizing
\[
{\cL'}(\mRh_1,\ldots,\mRh_K|\gl,\vs)=\sum_{k=1}^K N_k \left[ 
\log|\gl\mC_k^{-1}\va_k(\vs)\va_k\rH(\vs)+\mI|
+\tr\left(\left(\gl\mC_k^{-1}\va_k(\vs)\va_k\rH(\vs)+\mI\right)\mC_k^{-1}
\mRh_k\right)
\right].
\] 
After some algebraic manipulations described in the appendix, we obtain that
we have to minimize 
\begin{equation}
\label{simp_lik}
{\cL''}(\mRh_1,\ldots,\mRh_K|\gl,\vs)=\sum_{k=1}^K N_k\left[
\log(1+\gl\ga_k)-
\frac{\gl\gb_k}
{1+\gl\ga_k}
\right].
\end{equation}
where  
$\ga_k=\va_k\rH(\vs)\mC_k^{-1}\va_k(\vs)$ and 
$\gb_k=\va_k\rH(\vs)\mC_k^{-1}\mRh_k \mC_k^{-1}\va_k(\vs)$.
Taking the derivative of the left hand side with respect to $\gl$ yields
that the MLE of $\gl$ is given by solving the equation
\begin{equation}
\label{mle_lambda}
\frac{\partial \cL''}{\partial \gl}=\sum_{k=1}^K N_k\left[
\frac{\ga_k}{1+\gl\ga_k}-\frac{\gb_k}{1+\gl\ga_k}+\frac{\gl\ga_k\gb_k}
{(1+\gl\ga_k)^2}
\right]=0.
\end{equation}
Equation (\ref{mle_lambda}) is highly non-linear and hard to solve in the 
general case. We propose to simplify the problem by the following procedure. 
First estimate the source power using each of the covariance matrices 
separately and then average the estimates. Assuming that the ML estimate is 
consistent the above estimate will still be consistent. If we further 
assume that the statistical behavior of the
various estimates is approximately the same (i.e., the Fisher information 
based on each time observation is almost constant) then the estimator is
still efficient. This is formulated in the following lemma.

\blm
Assume that $\cL(\vx|\gl)=\sum_{k=1}^K \cL_k(\vx_k|\gl)$. Let 
$\glh_k=\arg \max_{\gl} \cL_k(\vx_k|\gl)$ be the MLE based on the $k$'th block 
of data. Assume that all $J_k$ are equal say $J_k(\gl)=J$, where
$J_k$ is the Fisher information for $\gl_k$ based on the data 
$\vx_k$, and the likelihood $\cL_k(\vx|\gl)$. 
Then ${\bar \gl}=\frac{1}{K}\sum_{k=1}^K \glh_k$ is an asymptotically 
efficient estimator of $\gl$.
\elm
Proof: The Fisher information of the overall likelihood is given by
$\sum_{k=1}^K J_k$. Thus the CRB on estimating $\gl$ is given by
\[
var(\glh) \ge \frac{1}{\sum_{k=1}^K J_k}=\frac{1}{KJ}.
\]
Since the MLE is asymptotically efficient (in the total number of samples 
$\sum_{k} N_k$) its
asymptotic variance achieves the CRB.
On the other hand if we estimate $\gl$ based on the $k$'th block we obtain
(by its asymptotic efficiency in $N_k$) that it has a variance given by
\[
var(\glh_k) = \frac{1}{J_k}.
\]
Now combining the various estimates we obtain:
\[
var({\bar \gl})=\frac{1}{KJ}.
\]
\qed
By the inequality of the harmonic and arithmetic mean we know that if
the Fisher information is different at some time instances, then the averaged
estimator has a larger variance, but the difference depends on the variation 
of the $J_k$'s. However this degradation in performance
gives us a large computational saving.
By the additivity of the derivative we obtain from (\ref{mle_lambda}) that 
for each $k$ the MLE 
$\glh_k$ based on $\mRh_k$ is given by 
\begin{equation}
\label{est_lmk}
\glh_k=
\frac{\va_k\rH(\vs)\left(\mC_k^{-1}\mRh_k \mC_k^{-1}-\mC_k^{-1}\right)
\va_k(\vs)}
{(\va_k\rH(\vs)\mC_k^{-1}\va_k(\vs))^2}
\end{equation}
Equivalently this can be written as
\begin{equation}
\glh_k=
\frac{\va_k\rH(\vs)\mC_k^{-1}\left(\mRh_k-\mC_k\right) \mC_k^{-1}
\va_k(\vs)}
{(\va_k\rH(\vs)\mC_k^{-1}\va_k(\vs))^2}
\end{equation}
and hence 
\begin{equation}
\label{AML_lambda}
{\bar \gl}=\frac{1}{\sum_{k=1}^K N_k}\sum_{k=1}^K N_k\left[
\frac{\va_k\rH(\vs)
\mC_k^{-1}\left(\mRh_k-\mC_k\right) \mC_k^{-1}\va_k(\vs)}
{(\va_k\rH(\vs)\mC_k^{-1}\va_k(\vs))^2}
\right].
\end{equation}

Now that we have this approximate ML (AML) estimate of $\gl$ we can plug it 
into the likelihood function and obtain that we have to maximize over the 
field of view the following:
\begin{equation}
\label{AML_theta}
{\hat \vs}=\arg \max_{\vs} \sum_{k=1}^K\log\left(|1+{\bar \gl}(\vs)\ga_k(s)|
\right)-
\frac{{\bar \gl}(\vs)\gb_k(\vs)}{1+{\bar \gl}(\vs)\ga_k(\vs)}.
\end{equation}

The proposed  coordinate-wise AML estimator is now summarized in table 
\ref{AML_table}.
\footnote{The stopping condition at step 4 
can be tested either using a $\chi^2$ statistic or by comparing the level of 
the extracted point source to the noise level.}
\footnote{The final MLE focusing operation can use the same update equations
for an alternating coordinate maximization. In this case we use the matrix 
inversion lemma twice: First we add the contribution of the last
estimated coordinate to $\mC_k$,
and then we subtract from $\mC_k$ the estimated contribution of the 
coordinate along which we would like to optimize.}

Alternatively to substituting back into the likelihood we can try to find the 
direction which maximizes the received power. It is interesting to observe 
that the CLEAN algorithm can be derived as an approximation to this power 
maximizing estimator. To that end  
maximize the power received from direction $\vs$, i.e., 
\begin{equation}
{\hat \vs}=\argmax_{\vs}{\bar \gl}(\vs).
\end{equation}
Using (\ref{AML_lambda}) we obtain:
\begin{equation}
\label{power_lambda}
{\hat \vs}=\arg \max_{\vs}\frac{1}{\sum_{k=1}^K N_k}\sum_{k=1}^K N_k\left[
\frac{\va_k\rH(\vs)
\mC_k^{-1}\left(\mRh_k-\mC_k\right) \mC_k^{-1}\va_k(\vs)}
{(\va_k\rH(\vs)\mC_k^{-1}\va_k(\vs))^2}
\right].
\end{equation}
From (\ref{self_cal_model}) we obtain that 
$\mC_k= \mAt_k \mB \mAt_k\rH + \sigma^2 \mI$, where $\mAt$ contains all 
the array responses towards the previously estimated astronomical sources.
Assuming now that the power of the astronomical sources is negligible compared
to the noise power (this is reasonable in many circumstances) and noting that 
that $\|\va(\vs)\|^2=p$ for every $\vs$ by our normalization,
we obtain that
$\mC_k \approx \gs^2\mI$ and that 
(\ref{power_lambda}) simplifies to (\ref{CLEANbf}).
An iterative application of (\ref{CLEANbf}) is exactly the operation of the 
CLEAN beam removing technique.

\section{Conclusions}
In this paper we have presented a parametric approach to radio-astronomical 
image formation. We have used this approach to adapt some known spectral 
estimators to the astronomical image formation problem. We analyzed the 
effect of interference suppression on imaging and  proposed the necessary 
changes to the imaging step in order 
to accommodate the spatial filtering pre-processing. A new 
AML algorithm for deconvolution has been presented, and the 
CLEAN algorithm was derived by approximating the ML power estimates. Finally 
we  have presented some simulated images demonstrating some of the ideas 
presented, and demonstrating the possible advantages in the parametric
approach that leads to improved resolution. 

The work shows that the design of a new radio-telescope can (and probably
should) use phased-array techniques to mitigate RFI, but the imaging software 
will have to be changed accordingly.

In this paper we have only analyzed LS based deconvolution. In extension of 
this work \cite{leshem99u1} we analyze the MEM image formation, in the context
of adaptive interference suppression, and propose suitable changes to the 
imaging. The possible applications of this work spans beyond the field of 
radio-astronomy. One possible example is ISAR imaging in the presence of 
strong interferers.

\section*{Acknowledgements}

    We would like to thank Ed Deprettere and our project partners at NFRA, 
    especially A.~van Ardenne, A.J.~Boonstra, G. van Diepen, P.~Friedman, 
    A.~Kokkeler, J.~Noordam, and G.~Schoonderbeek, for the very useful 
    collaboration.

\appendix
\section{Appendix A}					\label{sec:appA}

In this appendix we derive formula (\ref{simp_lik}).
To simplify the derivation we omit the explicit dependence on $\vs$.
Let $\mR_k=\mC_k\left[\mI+\gl\mC_k^{-1}\va_k\va_k\rH\right]$.
hence $\mR_k^{-1}=\left[\mI+\gl\mC_k^{-1}\va_k\va_k\rH\right]^{-1}\mC_k^{-1}$.
To compute $\left[\mI+\gl\mC_k^{-1}\va_k\va_k\rH\right]^{-1}$, we use the 
matrix inversion lemma obtaining:
\[
\left[\mI+\gl\mC_k^{-1}\va_k\va_k\rH\right]^{-1}=\mI-
\frac{\gl}{1+\gl\va_k\rH\mC_k^{-1}\va_k}\mC_k^{-1}\va_k\va_k\rH.
\]
Using $\tr(\va\vb\rH)=\vb\rH\va$, we obtain:
\[
\tr\left(\mR_k^{-1}\mRh_k\right)=\tr\left(\mC_k^{-1}\mRh_k\right)-
\frac{\gl\va_k\rH\mC_k^{-1}\mRh_k\mC_k^{-1}\va_k}{1+\gl\va_k\rH\mC_k^{-1}\va_k}.
\]
Since $\tr\left(\mC_k^{-1}\mRh_k\right)$ is independent of the parameters
it can be omitted from the maximization.

To evaluate $\log\left(|\mI+\gl\mC_k^{-1}\va_k\va_k\rH| \right)$ note that
the vector $\mC_k^{-1}\va_k$ is an eigenvalue of the rank one matrix 
$\gl\mC_k^{-1}\va_k\va_k\rH$ with eigenvalue $\gl\va_k\rH\mC_k^{-1}\va_k$.
Therefore it is an eigenvector of  $\mI+\gl\mC_k^{-1}\va_k\va_k\rH$ with 
eigenvalue
$1+\gl\va_k\rH\mC_k^{-1}\va_k$.
All the other eigenvalues of $\mI+\gl\mC_k^{-1}\va_k\va_k\rH$, are 1. Hence
since the determinant is the product of the eigenvalues 
$|\mI+\gl\mC_k^{-1}\va_k\va_k\rH|=1+\gl\va_k\rH\mC_k^{-1}\va_k$.
Substituting into $\cL$ we obtain (\ref{simp_lik}).

\section{Appendix B}					\label{sec:appB}

In this appendix we present the gradient
of the log-likelihood and the Fisher information matrix for 
the likelihood function given in equations (\ref{cov_model}),(\ref{eq:lik}).
This gives a possibility of obtaining rapidly convergent ML estimation,
given a good initialization.
We will not give the derivation in detail as it is rather standard. Similar 
derivation for the perfectly calibrated array model case can be found e.g., in 
\cite{sheinvald97phd}.

Let the parameter vector be
\[
\vpsi=\left[ 
\vl,\vm,\vgl,\vgrg_1,\ldots,\vgrg_K
\right].
\]
where 
$\vl=[l_1,\ldots,l_d]$ is the vector of $l$  coordinate of the astronomical
point sources, $\vm=[m_1,\ldots,m_d]$ is the vector of $m$  coordinate of the 
astronomical point sources (and $\vs_i=(l_i,m_i)$) are the location 
coordinates of the $i$'th astronomical source, 
$\vgl=[\gl_1,\ldots,\gl_d]$ is the vector of brightness of the sources,
and $\vgrg_k$ are the calibration parameters at the $k$'th observation epoch.
The score function is given by
\begin{equation}
\frac{\partial\cL}{\partial \vpsi}=
-\sum_{k=1}^K N_k \frac{\partial \hbox{vec}(\mR_k)}{\partial \vpsi}
\left(\mR_k^{-T} \otimes \mR_k^{-1} \right)
\hbox{vec}(\mRh_k-\mR_k).
\end{equation}
and 
\begin{equation}
\mJ(\vpsi)=\sum_{k=1}^K N_k \left(
\frac{\partial \hbox{vec}(\mR_k)}{\partial \vpsi}
\right)\rH
\left(\mR_k^{-T} \otimes \mR_k^{-1} \right) \frac{\partial 
\hbox{vec}(\mR_k)}{\partial \vpsi}
\end{equation}
We now turn to compute 
$\frac{\partial \hbox{vec}(\mR_k)}{\partial \vpsi}$. 
Equation (\ref{cov_model}) can be rewritten as:
\begin{equation}
\hbox{vec}(\mR_k)=\left[
\va_k\rH(\vs_1)\mgG_k\rH\otimes \mgG_k\va_k(\vs_1),\ldots,
\va_k\rH(\vs_d)\mgG_k\rH\otimes \mgG_k\va_k(\vs_d)
\right]\vgl
\end{equation}
Hence we obtain
\begin{equation}
\frac{\partial 
\hbox{vec}(\mR_k)}{\partial \vgl}=
\left[
\va_k\rH(\vs_1)\mgG_k\rH\otimes \mgG_k\va_k(\vs_1),\ldots,
\va_k\rH(\vs_d)\mgG_k\rH\otimes \mgG_k\va_k(\vs_d)
\right]
\end{equation}
and similarly
\begin{equation}
\frac{\partial 
\hbox{vec}(\mR_k)}{\partial \vl}=
\left[
\begin{array}{c}
\gl_1\left(\left(\frac{\partial \va_k}{\partial l}\right)\rH(\vs_1)\mgG_k\rH
\otimes \mgG_k\va_k(\vs_1)+
\va_k\rH(\vs_1)\mgG_k\rH\otimes \mgG_k \frac{\partial \va_k}{\partial l}(\vs_1)
\right)^T \\
\vdots \\
\gl_d\left(\left(\frac{\partial \va_k}{\partial l}\right)\rH(\vs_d)\mgG_k\rH\otimes 
\mgG_k\va_k(\vs_d)+
\va_k\rH(\vs_d)\mgG_k\rH\otimes \mgG_k \frac{\partial \va_k}{\partial l}(\vs_d)
\right)^T
\end{array}
\right]^T
\end{equation}
\begin{equation}
\frac{\partial 
\hbox{vec}(\mR_k)}{\partial \vm}=
\left[
\begin{array}{c}
\gl_1\left(\left(\frac{\partial \va_k}{\partial m}\right)\rH(\vs_1)\mgG_k\rH
\otimes \mgG_k\va_k(\vs_1)+
\va_k\rH(\vs_1)\mgG_k\rH\otimes \mgG_k \frac{\partial \va_k}{\partial m}(\vs_1)
\right)^T \\
\vdots \\
\gl_d\left(\left(\frac{\partial \va_k}{\partial m}\right)\rH(\vs_d)\mgG_k\rH\otimes 
\mgG_k\va_k(\vs_d)+
\va_k\rH(\vs_d)\mgG_k\rH\otimes \mgG_k \frac{\partial \va_k}{\partial m}(\vs_d)
\right)^T
\end{array}
\right]^T
\end{equation}
Finally the derivatives with respect to the calibration parameters
are given by:
\begin{equation}
\frac{\partial (\mR_k)_{ij}}{\grg_{n,k}}=\gd_{i,n}\sum_{l=1}^d a_i(\vs_l) 
\grg_{j,k}^*a_j^*(\vs_l)+\gd_{j,n}\sum_{l=1}^d \grg_{i,k} a_i(\vs_l) a_j^*(\vs_l).
\end{equation}
where $\gd_{i,k}$ is Kronecker's delta, and $\grg_{n,k}$ are given by 
(\ref{def_Gammak}).


\newpage
\section*{Biography}
{\bf Amir Leshem}(M' 98) was born in Israel in 1966. He received the B.Sc.
(cum laude) in mathematics and physics, the M.Sc. (cum laude) in mathematics, 
and the Ph.D. in mathematics all from the Hebrew University, Jerusalem, Israel,
in 1986,1990 and 1997 respectively.
From 1984 to 1991 he served in the IDF.
From 1990 to 1997 he was a researcher with RAFAEL signal processing
center. From 1992 to 1997 he was also a teaching assistant at the
Institute of mathematics, Hebrew University, Jerusalem.
Since 1998 he is with Faculty of Information Technology and Systems,
Delft university of technology, The Netherlands, working on algorithms for
the reduction of electromagnetic interference in radio-astronomical 
observations, radio astronomical imaging techniques and signal processing for 
communication. 

His main research interests include array and statistical signal processing, 
radio-astronomical imaging methods, set theory, logic and foundations of 
mathematics.

    {\bf Alle-Jan van der Veen} (S'87, M'94) was born in The Netherlands in
        1966.  He graduated (cum laude) from Delft University of Technology,
        Department of Electrical Engineering, in 1988, and received the
        Ph.D.\ degree (cum laude) from the same institute in 1993.
        Throughout 1994, he was a postdoctoral scholar at Stanford
        University, and he has held visiting researcher positions at
        Stanford, Chalmers University, and ENST Paris.  At present, he
        is associate professor in the Signal Processing group of DIMES,
        Delft University of Technology.

\newpage
\listoftables
\listoffigures

\newpage

\begin{table}
\caption{The CLEAN algorithm} 
\label{tab:CLEAN}
\begin{center}
    \fbox{\quad
    \parbox{.45\textwidth}{
        $l=0$\\
	while $I_D$ is not noise-like:
	\[\left[
	\bea{lcl}
	      \multicolumn{3}{l}{
		  (\ell_l,m_l) = \argmax I_D(\ell,m)  } \\
	      \lambda_l &=& I_D(\ell_l,m_l)/B_0(0,0)\\
	      I_D  &:=& I_D - \gamma \lambda_l B_0(\ell-\ell_l,m-m_l)\\
	      l &=& l + 1
	\ena\right.
        \]
	$I = I_D + \sum_l\, \gamma \lambda_l B_{synth}(\ell-\ell_l,m-m_l)$
    }}
\end{center}
\end{table}

\begin{table}
    \caption{The CLEAN algorithm with spatial filtering}
    \label{tab:clean_filter}
\begin{center}
    \fbox{\quad
    \parbox{.65\textwidth}{
	\mbox{Compute $\mRh'$ using (\ref{mRhpdef})}\\
	$I_D'(\vs) = \sum_{k=1}^K \va_k\rH(\vs) \mRh'_k \va_k(\vs) $\\
        $l=0$\\
	while $I_D'$ is not noise-like:
	\[
	  \left[ \bea{lcl}
	      \bs_l &=& \argmax I_D'(\vs) \\
	      \multicolumn{3}{l}{
\mbox{Compute $B(\bs,\bs_l)$ using (\ref{defB})}

}\\
	      \lambda_l &=& I_D(\bs_l)/B(\vs_l,\vs_l)\\
	      I_D'(\vs)  &:=& I_D'(\vs) \;-\; \gamma \lambda_l B(\vs,\vs_l)\\
	      l &=& l + 1
	   \ena\right.
        \]
	$I = I_D' + \sum_l\, \gamma \lambda_l B_{synth}(\vs-\vs_l)$
    }}
\end{center}
\end{table}

\begin{table}
    \caption{Iterative approximate maximum likelihood}
    \label{AML_table}
\begin{center}
    \fbox{\quad
    \parbox{.65\textwidth}{
\begin{enumerate}
\item Set the data covariance to be $\left\{\mRh_k^{(0)} \ :\  k=1,\ldots,K \right\}$. 
\item Set the noise covariance matrix to be $\mC_k^{(0)}=\gs^2\mI$. 
\item Let $\mgD_k^{(0)}=\mRh_k^{(0)}-\mC_k^{(0)}$.
\item Until the residual is noise like 
perform the following:
    \begin{enumerate}
	\item Estimate the direction of a point source $\vsh_l$ using 
             (\ref{AML_theta}).
         \item Estimate $\glh_l$ using (\ref{AML_lambda}).
         \item Let $\mC_k^{(l+1)}=\mC_k^{(l)}+\grg \glh_l \va(\vsh_l) 
               \va(\vsh_l)^h$ 
         \item Update $\mgD_k^{(l+1)}=\mgD_k^{(l)}-\mC_k^{(l+1)}$. 
         \item Update $\left(\mC_k^{(l+1)}\right)^{-1}$ using the matrix 
               inversion rank 1 update formula.
         \item $l:=l+1$.
    \end{enumerate}
\item Use the estimated components to initialize a MLE search.
\item Reconstruct the image using the estimated directions, estimated powers 
      and an ideal beam-shape.
\end{enumerate}
}}
\end{center}
\end{table}

\newpage


\newpage

\newpage

\begin{figure}
\begin{center}
    \mbox{\psfig{figure=ja_clean61.epsi,width=0.5\textwidth}}

    \mbox{\psfig{figure=ja_dirty61.epsi,width=0.5\textwidth}}

    \mbox{\psfig{figure=ja_filter61.epsi,width=0.5\textwidth}}
\end{center}
    \caption{Dirty images of four point sources. $(a)$ no interference;
	$(b)$ unsuppressed interference ($\INR = 5$~dB); $(c)$ after spatial
	filtering.}
    \label{Fig:spatialclean}
    \label{Fig:spatialinterf}
    \label{Fig:spatialfilter}
\end{figure}

\end{document}